\documentclass[12pt]{article}
\usepackage{latexsym}
\usepackage{amssymb}
\usepackage{amsmath}
\usepackage{graphicx}
\usepackage{enumitem}
\usepackage{bbold}
\usepackage{slashed}
\usepackage{hyperref}

\def\be{\begin{equation}}
\def\ee{\end{equation}}
\def\bear{\begin{eqnarray}}
\def\eear{\end{eqnarray}}
\def\nn{\nonumber}

\def\g{\gamma}
\def\d{\delta}

\def\r{\rho}

\def\Tr{{\rm Tr}}

\def\dd{\mbox{d}}

\def\d{\delta}
\def\D{\Delta}

\def\g{\gamma}

\def\e{\epsilon}

\def\F{\Phi}

\def\l{\lambda}

\def\r{\rho}

\newcommand{\sm}[1]{\mbox{\scriptsize #1}}
\newcommand{\tn}[1]{\mbox{\tiny #1}}
\renewcommand{\@}[1]{\sqrt{#1}}
\renewcommand{\le}[1]{\label{#1}\end{eqnarray}}
\newcommand{\bea}{\begin{eqnarray}}
\newcommand{\eea}{\end{eqnarray}}

\newcommand{\eq}[1]{(\ref{#1})}
\def\nn{\nonumber\\}

\def\ffract#1#2{\raise .35 em\hbox{$\scriptstyle#1$}\kern-.25em/
\kern-.2em\lower .22 em \hbox{$\scriptstyle#2$}}

\begin{document}
\pagestyle{empty}

\centerline{{\Large \bf Interpreting Theories without a Spacetime }}
\vskip 1truecm

\begin{center}
{\large Sebastian De Haro$^1$ and Henk W.~De Regt$^2$}\\
\vskip .7truecm
$^1${\it Trinity College, Cambridge, CB2 1TQ, United Kingdom}\\
{\it Department of History and Philosophy of Science, University of Cambridge\\
Free School Lane, Cambridge CB2 3RH, United Kingdom\\
Vossius Center for History of Humanities and Sciences, University of Amsterdam}\\
\vskip0.5truecm
$^2${\it Faculty of Philosophy, VU University Amsterdam}\\
{\it De Boelelaan 1105, 1081 HV Amsterdam, The Netherlands}

\vskip .5truecm
{\tt sd696@cam.ac.uk,~~h.w.de.regt@vu.nl}
\vskip .5truecm 
\end{center}

\vskip 9truecm

\begin{center}
\textbf{\large \bf Abstract}
\end{center}

In this paper we have two aims: first, to draw attention to the close connexion between interpretation and scientific understanding; second, to give a detailed account of how theories without a spacetime can be interpreted, and so of how they can be understood.

In order to do so, we of course need an account of what is meant by a theory `without a spacetime': which we also provide in this paper.

We describe three tools, used by physicists, aimed at constructing interpretations which are adequate for the goal of understanding. We analyse examples from high-energy physics illustrating how physicists use these tools to construct interpretations and thereby attain understanding. The examples are: the 't Hooft approximation of gauge theories, random matrix models, causal sets, loop quantum gravity, and group field theory.

\newpage
\pagestyle{plain}

\tableofcontents

\newpage

\section*{Introduction}\label{intro}
\addcontentsline{toc}{section}{Introduction}

This paper has two aims: First, to briefly expound the close connexion between interpretation and understanding (section \ref{titu}): which has not received prior attention in the literature on scientific understanding. While the close connexion between understanding and interpretation has been acknowledged in the social sciences and humanities, it has been largely ignored in debates regarding understanding in the natural sciences.\footnote{Interpretation in the natural sciences is discussed in e.g.~van Fraassen and Sigman (1993) and Faye (2014).} But we will see that interpretation is in fact essential in promoting understanding in modern theories of high-energy physics, so that it is worth exploring the connexion. Second, to catalogue and analyse the ways in which interpretations are constructed, and so the ways in which understanding can be had, in physical theories that do not contain a spacetime (sections \ref{3tools}, \ref{viseff}, and \ref{nost}). In a companion paper (De Haro and De Regt (2018)), we analyse recent debates in the philosophical literature on the merits of theories without spacetime and we argue that, although spacetime and visualisation are desirable tools for interpretation and understanding, they are not necessary.\\

Current theoretical research in quantum gravity suggests that space and time may not have the fundamental status that has traditionally been bestowed upon them. On the contrary, in recently proposed theories of quantum gravity the fundamental notions of space and time seem to be absent; they are supposed to be replaced by alternative, non-spatiotemporal structures. An important question which arises, if such theories without spacetime are considered a live option, is how they should be interpreted. This is especially pressing for scientific realists, who regard theories as (ideally) true descriptions of reality and who will therefore ask: How can the real world possibly be the way a theory without spacetime says it is? Do space and time not exist at the fundamental level of reality? But the question is also urgent for those who do not regard themselves as realists. Constructive empiricists, for example, will need to know how theories without spacetime can yield empirically adequate models of the observable physical world, and moreover, what a literal construal of the theory would look like.\footnote{Although constructive empiricists withhold belief in the truth of the theory, they do construe the theory literally and can use the literal interpretation of the theory for heuristic or explanatory purposes.} Interpretation is thus crucial for answering questions about the representational virtues of scientific theories. But representation is not the only aim of science. Other goals are prediction, explanation, and understanding: and achieving these goals requires interpretation as well.

Understanding of phenomena, for example, has traditionally been regarded as requiring theories that are formulated in a spacetime framework, because this is a necessary condition for visualisation, which is regarded as an important---even indispensable---tool for understanding. So, how can physical theories in which space and time are fundamentally absent provide scientific understanding, if at all? Again, the answer to this question depends on how theories without spacetime can or should be interpreted. In a companion paper (De Haro and De Regt (2018)) we analyse recent debates in the philosophical literature on the merits of theories without spacetime. It appears that these debates reflect disagreements about whether or not such theories are intelligible and whether they can provide understanding.\footnote{A related problem is the problem of `empirical incoherence'. For a discussion, see the special issue of {\it Studies in History and Philosophy of Modern Physics}, edited by Huggett and W\"uthrich (2013).} We argue that interpretation in terms of a spacetime framework is not necessary for understanding, since visualisation is only one out of many possible tools for rendering theories intelligible. Understanding may also be achieved by other means, with other tools. 

In the present paper we develop specific tools---visual as well as non-visual ones---that are particularly useful for the interpretation and understanding of theories without spacetime. Our analysis focuses on interpretation, but we also discuss its connexion with understanding. We analyse various contemporary theories of quantum gravity without a spacetime, and show how the alternative tools can be applied to render these theories intelligible. \\

The outline of the paper is as follows. In section \ref{titu}, we discuss the connexion between interpretation and understanding: i.e.~what it means to interpret a physical theory, and how interpretations function as tools for achieving scientific understanding of phenomena. Section \ref{3tools} presents three tools for interpreting theories without spacetime: (1) approximations, (2) similarities, and (3) internal criteria. In the subsequent two sections, these tools are used to interpret a variety of quantum gravity theories. Thus section \ref{viseff} focuses on theories for which visualisation via effective spacetimes is possible ('t Hooft's planar diagrams and random matrix models), while section \ref{nost} deals with theories with interpretations that either completely resist visualisation, or do not require it for the problem at hand, so that they are interpreted in a non-visual manner (causal sets,  loop quantum gravity and spin foams, group field theory). Section \ref{conclusion} concludes.

\section{Interpretation and understanding}\label{titu}

In this section, we discuss the relation between interpretation and understanding. We lay out, in section \ref{wiin}, our preferred conception of interpretation. Then, in section \ref{iau}, we summarise our preferred theory of understanding, and note that interpretation is a precondition for understanding.

\subsection{What is an interpretation?}\label{wiin}

In this section, we sketch some of the recent literature on interpretation, and place our preferred conception of interpretation against this background. In this literature, understanding is seen as one of the aims of interpretation.

If we want to know what a scientific theory says about the world, or if we want to use a theory for predicting, explaining or understanding empirical phenomena, we need to {\it interpret} that theory. Abstract theories can only be put to concrete use if they are interpreted. As Bas van Fraassen (1989, p.~226) states in {\it Laws and Symmetry}, ``any question about content [of a scientific theory] is, in actuality, met with an interpretation.'' So, interpretation is crucial, but what is it and what does it involve? According to the standard view---or the ``ideal view'', as Hoefer and Smeenk (2016) call it---the interpretation of a physical theory ``should characterize the physical possibilities allowed by the theory as well as specifying how the mathematical structures used by the theory acquire empirical content'' (Hoefer and Smeenk 2016, p.~118). This standard account of interpretation is widely accepted and implicit in many philosophical accounts of science. A recent statement and critical analysis of it can be found in Laura Ruetsche's {\it Interpreting Quantum Theories} (2011). In line with the characterisations given above, she describes the standard account as asserting that ``the content of a theory is given by the set of worlds of which that theory is true'' (p.~6), and ``to interpret a physical theory is to characterize the worlds possible according to that theory'' (p.~7). 

Ruetsche (2011, pp.~3-4) observes that this standard view is typically associated with an ``ideal of pristine interpretation'', which sees the business of the interpreter as a ``lofty'' affair that is only concerned with the general question of which worlds are possible according to the theory, and not with the application of the theory to actual physical systems in the world. On this ``pristine'' ideal, the question of which world is actually realised is merely a contingent matter and (therefore) philosophically uninteresting. While she does not reject the standard account itself, Ruetsche criticises the associated ideal of pristine interpretation and argues for ``a less principled and more pragmatic approach to interpreting physical theories, one which allows `geographical' considerations to influence theoretical content, and so allows the same theory to receive different interpretations in different contexts'' (p.~4).\\ 

In this paper, we will adopt a conception of interpretation proposed by one of us, which captures both elements mentioned by Ruetsche (for more details, see De Haro (2016) and De Haro and Butterfield (2017)). On this conception, an interpretation is characterised as follows:\\

{\bf Interpretation:} an interpretation of a theory, $T$, is a (partial) map, $i$, preserving appropriate structure, from the theory to a domain of application, $D$, within a possible world, $W$, i.e.~a map $i: T \rightarrow D$.

There are different kinds of interpretations, tuned to different purposes (see below). \\

It is worth clarifying, here at the outset, some possible misleading connotations of the above conception of interpretation, which we reject:\footnote{We thank an anonymous referee for suggesting to clarify these two points.}\\
\\
(a)~~{\it Not a model-theoretic conception of interpretation:} While our notion of interpretation bears some resemblance with the model-theoretic notion of interpretation (of which Ruetsche presents the somewhat bleak version that she calls the `pristine ideal'), ours is in fact not a model-theoretic conception. According to the latter, a theory is often identified with an entire class of models, which are each, individually, a consistent interpretation of the language of the theory (alternatively, an interpretation is the set of possible worlds in which the theory is true). On this view, a family of models constitutes the truth conditions for the theory, that is, all the possible interpretations that provide the theory with a truth value. Our notion differs from this in two ways: 

(i)~\,The domain of application, $D$, should not be confused with a `model' of the theory, in the formal sense. Rather, the domain of application, $D$, is a part of the world: it is not ``more theory'', but rather the range of the map $i$ is to be straightforwardly interpreted as being ``in the world''. And so, depending on the kind of interpretation one is considering,\footnote{See `Understanding as an aim of interpretation', later in this section.} the items in this domain can be interpreted ontologically, or as observational or experimental procedures or activities carried out by scientists, as material artefacts constructed by scientists, etc. In other words, the elements of the domain of application, $D$, are not formal objects---even if it is sometimes useful to present them as formalised, specifically: as elements of sets, and as relations between the elements. 

(ii)~~Our conception of interpretation does not require truth-values or truth-conditions, because most interpretations we consider (certainly the ones in this paper) are not literally `true'. In fact, our conception of interpretation does not use the notion of `truth'. Rather, the relation between the formal theory and the world is much looser: it is not specified by `truth-values', but it is explicated via the much more flexible notion of a partial map: which of course need not be (and in general is not!) bijective, nor even injective. \\
\\
(b)~~{\it Applicability beyond the mathematical aspects of theories:} A partial map is a flexible notion,\footnote{Although flexibility can also have some disadvantages, we submit that our conception does not have these disadvantages, and that the kind of flexibility discussed here is something we want: see our comments, below, on the logical weakness of the conception of interpretation, and on the aims of an interpretation. Our conception of interpretation is flexible, but appropriate strengthenings can be adopted: depending on the aim of the interpretation.} which maps not only between mathematical objects, but also between other objects: and this is something we want. Indeed, the map has a domain (the bare theory, $T$) and a range (the domain of application, $D$), which are both sets. As already mentioned under (a), the domain of application, $D$, can contain such diverse items as natural entities, numbers on a scale, results of experiments, and even human actions (including methodological procedures: often as relations between elements in the domain of application). As such, this conception of interpretation is closer to the ``mediator'' conception of a model, where a model is seen as mediating between the theory and the phenomena. A map can indeed provide such a mediation, between the theory (or a particular instantiation, or sector, of the theory) and the domain of application, which often consists of the particular phenomena the theory aims to describe. And the map can be as complex as you want. These kinds of contextual details belong to interpretations that are extensions (see the paragraph just below). Similar remarks can be made about the theory, $T$, which need not be a formal theory, but admits diverse degrees of formalisation, adapted to the branch of science in question. Needless to say: in this paper, we will concentrate on mathematically formulated theories, but we do not see any bar to applying this conception of interpretation far beyond the domain discussed in this paper.\\

For the purposes of this paper, the central distinction will be between two different kinds of interpretation maps: an intension map and an extension map, which are best characterised using the Frege-Carnap-Lewis framework for semantics (Frege 1892, Carnap 1947, Lewis 1970). An {\it intension} is, roughly, the linguistic meaning of a term in a theory: expressed quantitatively (if the theory is quantitative), but in general terms. And an {\it extension} is, roughly, the worldly reference of the term, relative to a single possible world, within a specific context of experiment or description, including all the contingent (and, admittedly, messy!) details involving the application of a theory to a concrete physical situation. 

Notice that the distinction between intension and extension is not tied up with a particular metaphysical position, but simply relies on an analysis of how terms refer in ordinary language: and similarly for scientific theories. Intensions map to an idealised world of meanings, independently of particular instances (one can also say---it holds across all possible worlds described by the theory). Extensions, on the other hand, are relative also to all other features that, together, determine reference: such as, for example, `the voltmeter which I have in my hand reads $220$ V'. This sentence will clearly have different meanings when uttered by different scientists, and even with different voltmeters because of measurement errors; and so, it is to be interpreted differently---i.e.~to receive a different extension---in different contexts.

An account of interpretation, and of its role in scientific theories, must take into account both kinds of interpretations, the intensional and the extensional: since both are used in scientific practice. 

The general conception of interpretation given above, as a structure-preserving partial map from the theory to a domain of application in the world, accounts for both kinds of cases: the differences being picked out by the properties of the domain of application, $D$, to which the interpretation maps, and the kind of world, $W$, within which that domain lies. Intension maps map to a domain within a world that is otherwise devoid of concrete features (such as: the person who is holding the thermometer, or the time of the day), because it abstracts from all possible worlds described by the theory; while extension maps map to to a domain within a possible world that possesses all the contingent features which are characteristic of, for example, a specific physical context (such as: when was the measurement made, at which location, etc.). 

It is important to clarify that the interpretation maps here discussed---as in the general case of intensions and extensions for linguistic items---need not be taken in a rigorous mathematical sense, but allow a degree of imprecision. The degree of precision obviously depends both on the accuracy of the theory, and on the level of detail of the domain. The more formal the theory, the more precise the interpretation map can be, and thus the more accurate the representation of reality that it can provide. But an interpretation is of course never merely formal (see the points (a) and (b) above). 

This conception is logically weak, in that there is very little that the existence of such a map requires. And there are various ways in which an interpretation can be made more restrictive (by imposing additional conditions on the map), and various kinds of interpretations that can be described---by further specifying the kinds of domains of application and worlds that are admissible  (cf.~De Haro (2016)).\\

{\it Understanding as an aim of interpretation.} Interpretation has various goals. Usually, as noted above, interpretation is related to realism: What are we committed to believe if we interpret a scientific theory realistically? In addition to representation, however, central aims of science are prediction, explanation, and understanding. Interpretation serves these goals as well; indeed, it is a precondition for achieving them. But it is not self-evident that these goals require the same considerations about interpretation as representational goals do. As to the aim of understanding, realists will typically be inclined to believe that there is a unique interpretation of the theory that provides understanding of the physical world by virtue of presenting the set of possible worlds of which the theory is true. Such realists may specify additional constraints; for example, they might require that the interpretation is in accordance with particular metaphysical presuppositions (a case in point being the fundamentally spatiotemporal nature of reality). Similar considerations will be endorsed for the predictive and the explanatory power of the theory. What such a priori considerations ignore, however, is the question of how an actual physical theory achieves these goals in practice. As Ruetsche (2011, p.~5) observes: 

``The lofty debate is conducted in terms that obscure how real theories possess the virtues they possess. It is often a theory {\it under an interpretation} that predicts, explains, and promotes understanding. To the disappointment of the realist, there may not be a single interpretation under which a given theory accomplishes all those things.'' 

Again, the framework of intensions and extensions accounts for the fact that a single theory may admit of many (and even mutually exclusive) interpretations. The kinds of interpretations, required for the other goals mentioned (prediction, explanation, and understanding), indeed do not need to be realistic, or accurate, scientific representation: but the kind of reference picked out by the interpretation may be tuned to the aims of prediction, explanation, and understanding. Thus we must carefully distinguish the {\it representation of a given system} (i.e.~the accurate scientific description of a {\it given} system, according to appropriate standards of empirical adequacy) from the theory's {\it reference}, with which the above conception of interpretation is concerned. And the question of representation still differs from the question of the truth of the theory: for both realists and constructive empiricists aim at representation, though their construals of these representations differ.

Let us give an example of an interpretation (in the above sense) that we will encounter in section \ref{causet}, and which does not give an accurate {\it representation} of a physical system by the theory, though it does successfully {\it refer} to a possible world, with the aim of promoting understanding. It will be an example of a theory of quantum gravity, interpreted by scientists, not in terms of physics, but in terms of family relations (including the notions of sibling, child, etc.). This interpretation invokes intensions, because the appeal is not to a particular family in the world but a generic, nameless one. Of course, these intensions are perfectly {\it in}compatible with the other, more common, intension, whose reference is to discrete spatiotemporal relations of succession that the theory purports to describe. But it does achieve its goal of promoting understanding, and it does satisfy the conception of interpretation which we gave earlier.

This example shows that this semi-formal framework for interpretation (also in De Haro (2016) and De Haro and Butterfield (2017)) is not committed to a single aim of science (say, a matter of representing reality),\footnote{For a discussion of this point in the context of quantum gravity, see Huggett and W\"uthrich (2013:~\S3), De Haro (2018:~\S3.1).} but simply analyses what scientists actually do when they interpret scientific theories. Thus we do not need to take a position in the debate, whether interpretations are ``really'' the lofty affair of pristine interpretation, or whether they are ``really'' the messy matter of coming to grips with a varied world that is difficult to grasp with a single interpreted theory. Rather, as the above example should show, the intension vs.~extension distinction cuts along a different edge: and it preserves for us two features that are key when interpreting theories, viz.~the conceptual and the concrete.\\

There are, so far, no direct experimental data about genuine quantum gravity effects.\footnote{For a discussion of the motivation for the quantum gravity programme, see e.g.~De Haro (2018:~\S4.1.1).} Consequently, all the interpretations that we will analyse in this paper, and in particular the three tools which we will develop, belong to the realm of intension. They are conceptual tools that physicists use, in the absence of adequate experimental or observational data, to build interpretations of their theories. 

\subsection{The contextual theory of understanding}\label{iau}

In this section, we review our preferred account of understanding, viz.~the contextual theory of understanding (from De Regt (2017)), and discuss the relation between interpretation and understanding further. 


Understanding of phenomena has traditionally been regarded as requiring theories that are formulated in a spacetime framework, because only such theories allow for visualisation, which is often regarded as an important, or even indispensable, tool for understanding. So, how can physical theories in which space and time are fundamentally absent provide scientific understanding, if at all? In a companion paper (De Haro and De Regt (2018)) we address this question by engaging with recent literature on the merits of recent theories without spacetime. It appears that these debates reflect disagreements about whether or not such theories are intelligible and whether they can provide understanding. Answers to these questions obviously depend on which conception of scientific understanding is adopted. In that paper, we argue that the contextual theory of scientific understanding (De Regt (2017)) is highly suitable for this task, because it is attuned to scientific practice, especially the practice of the physical sciences, and because it acknowledges that scientific understanding of phenomena is related to the intelligibility of theories, where the latter is a contextual matter in the sense that it depends on the skills of scientists and the available tools for understanding. On this theory of understanding, the question can be answered in the affirmative: theories without spacetime {\it can} be intelligible, and accordingly there is no principled obstacle for achieving scientific understanding with such theories.\\

The contextual theory of scientific understanding is based on the idea that scientists achieve understanding of a phenomenon $P$ if they construct an appropriate model of $P$ on the basis of a theory, $T$. More specifically, it contains the following criterion for scientific understanding (Criterion for Understanding Phenomena):\\

{\bf CUP:} A phenomenon $P$ is understood scientifically if and only if there is an explanation of $P$ that is based on an intelligible theory, $T$, and conforms to the basic epistemic values of empirical adequacy and internal consistency.\footnote{De Regt (2017:~p.~92). This is a revised formulation of CUP as presented in De Regt and Dieks (2005:~p.~150).}

The key term in this criterion is `intelligible': understanding phenomena requires an intelligible theory, where intelligibility is defined as:\\

{\bf Intelligibility:} the value that scientists attribute to the cluster of qualities of a theory that facilitate the use of the theory.\footnote{De Regt (2017:~p.~40). The original formulation contains the clause ``(in one or more of its representations)''. For our present purposes, we will not need to take this into account.}

A theoretical quality that is highly valued by scientists, past and present, is visualisability. Many scientists prefer visualisable theories because these are more tractable and easier to work with---in other words, visualisation is widely used as a tool for understanding (see De Regt (2014)). But the contextual theory of understanding does not entail that visualisability is a necessary condition for the intelligibility of scientific theories. It allows for alternative ways to render theories intelligible. Accordingly, although visualisation is an important tool for understanding, it is not indispensable. The fact that visualisation has proven to be a very effective tool in the past is not surprising because visual skills are so widespread and well-developed among humans. This explains why visualisation is a dominant strategy for achieving understanding, but it does not follow that it is a necessary condition for it. Hence, visualisation may be characterised as merely a `contingently dominant' tool for understanding. One aim of the present paper is to present alternative tools for interpreting theories without spacetime in such a way that they can still be rendered intelligible.\\

Why do scientific theories need to be intelligible to the scientists who use them? This is because scientists understand phenomena by constructing {\it models} of them, and this involves pragmatic judgments and decisions, since models do not follow straightforwardly from theories. An objective test for assessing the intelligibility of a theory (for a scientist in a particular context) is provided by the following Criterion for the Intelligibility of Theories:

{\bf CIT$_1$:} A scientific theory $T$ is intelligible for scientists (in context $C$) if they can recognise qualitatively characteristic consequences of $T$ without performing exact calculations.\footnote{De Regt (2017:~p.~102). CIT$_1$ is only one of many possible criteria (tests) for intelligibility, which may be termed CIT$_2$, CIT$_3$, CIT$_4$, etc.}

The basic idea behind CIT$_1$ is that the scientists have an `insight' into the workings of the theory and are accordingly able to use the theory for the construction of models of the phenomena, which satisfy the basic values of empirical adequacy and internal consistency. The resulting models provide the explanations that produce understanding of the phenomena in question. The conceptual tools---e.g. visualisation---that scientists may use to recognise qualitative consequences of $T$ are also the tools that will facilitate model construction.\\

The intelligible theory, $T$, is here to be understood as an interpreted theory, since a `bare' mathematical theory without an interpretation does not yield any predictions, let alone explanation or understanding. Only within an interpreted theory can one construct models of phenomena, in order to explain and predict. Recall, from section \ref{wiin} (b), the close connexion between our conception of interpretation and the idea of models, as mediators between theory and phenomena. While not every interpretation, $i$, necessarily involves models in this sense, many of them do (as our examples in sections \ref{viseff} and \ref{nost} will illustrate). More generally, interpretations proceed via the development of tools that allow scientists to recognise qualitatively characteristic consequences of the theory, $T$, without performing exact calculations (i.e.~a case of CIT$_1$). We will present these tools, for the theories without a spacetime described here, in section \ref{3tools}. The idea is that, once an interpretation, $i:T\rightarrow D$, is constructed, scientists can use the interpretation map, $i$, to make qualitative predictions. Thus, {\it interpretation is a precondition for intelligibility}, and different interpretations of $T$ may lead to different degrees of intelligibility. The task of interpreting a theory should therefore not be taken as a simple and straightforward matter, especially because it should not be assumed (as realists are wont to do) that there is a unique interpretation for every scientific theory. The development of an interpretation of a theory is the first step in rendering the theory intelligible and in achieving understanding of phenomena (present or, for quantum gravity, indeed future) with the theory. The choice for a particular interpretation will be guided by preferences for specific tools for understanding. Indeed, an interpretation is itself built using specific interpretative tools, which in turn are tools for understanding, because interpretation leads to understanding. 

Above, we discussed the construction of {\it models} as one way in which scientists understand phenomena, by linking them to the theory: models mediate between theory and phenomena, thus rendering theories intelligible. But also: models explain the phenomena, thus providing understanding of them. As such, models of phenomena (e.g.~the Bohr model of the atom and its interactions with photons) function as interpretations of theories, as claimed in section \ref{wiin} (b). 

In the next section, we will present three tools for interpreting theories without spacetime, and for rendering them intelligible. The application of these tools, in section \ref{viseff}, will show that even theories without spacetime can be interpreted in ways that allow for the use of visualisation. But the same tools can also be employed to construct non-visualisable interpretations that can still lead to intelligible physical theories, as will be shown in section \ref{nost}.

\section{Three Interpretative Tools}\label{3tools}

In this section, we identify three specific tools which are used by physicists to interpret theories, and in particular the theories without spacetime that concern us.\footnote{We use the term `tool' deliberately, to emphasise that our approach to interpretation and understanding is a pragmatic one. While the tools that we discuss may traditionally be regarded as `methods', we want to highlight how they function as conceptual tools that can be used to achieve desired ends, namely interpretations that render theories intelligible.}

Recall, from section \ref{wiin}, that we construe an interpretation as a map from the theory to an appropriate domain of application. The leading idea, in this section and the next, will be that there are three tools that physicists use in constructing (parts of the) interpretation of theories without a spacetime. The tools themselves contribute to the physicists' understanding of the theory, but they do so through the development of an interpretation. It is the interpretation that is guiding in understanding the theory, and in constructing models to explain particular phenomena. An interpretation is often, and traditionally, regarded as having a visual component (which is why visualisation is contingently dominant). But we will see that an interpretation {\it need not} be visual.



The three tools aim at developing interpretations of theories, constructing suitably explanatory models of phenomena, and recognising qualitative characteristic consequences of theories without performing exact calculations.

All the tools that we discuss in this paper are ``umbrella concepts'', realised by very different sorts of procedures that only bear some family resemblances with one another. Thus we will not attempt to provide, and we will not need, analytic definitions of the tools. It will suffice, for our purposes, to characterise the tools, and to point out the resemblances between the examples.\footnote{For a discussion of tools in quantum gravity, see De Haro (2018:~\S3).}

The first two tools relate {\it theories}: the first tool relates two theories via an approximative relation, or linkage; the second tool relates them in a looser way (via a relation of similarity, or analogy). On the other hand, the third tool compares terms within a {\it single theory} (again, through similarity or analogy). We will also give a variant of the first tool, in which the interpretive arrows are reverted. 

\subsection*{(Approximation) Use of approximative relations between theories}
\addcontentsline{toc}{subsection}{(Approximation)~~Use of approximative relations between theories}

It is common practice in physics to study theories in specific regimes of parameters, through various approximations or by taking limits. For example, in quantum field theory one studies the semi-classical limit (of a small value of the coupling constant), in which the theory reproduces the classical results, up to small corrections. In statistical mechanics, one studies the (unphysical) limit in which the number of particles and the volume of the system go to infinity, thus reproducing thermodynamics.\footnote{None of what we will do here hinges on a mathematical construal of the word `limit': indeed, in the two previous examples, it is widely understood that the relation between the physical systems need not (cannot) be that of a strict mathematical limit. This is indeed one of the main points of Wilson's renormalisation programme in condensed matter and quantum field theory: that cutoffs, which are introduced in order to regularise and renormalise theories, need not be `sent to zero', since they may in fact have a physical significance. The same is true about the `thermodynamic limit': a tea kettle of course boils despite its containing only a finite number of water molecules. Indeed, in many cases of interest, the limits do not even exist.}

So, the use of what we here call (Approximation) will include a diverse spectrum of such, not mutually exclusive, procedures and linkages between theories: 

(i) {\it Mathematical limits}, in which a variable is taken to some special value: as in a sequence in which a continuous or a discrete variable is taken to e.g.~zero (and the variable can be a parameter in an expression, model or theory). 

(ii) {\it Comparing different physical situations or systems}, e.g.~as when a large value of the action is considered (in addition to taking $\hbar\rightarrow0$) in the transition from quantum to classical mechanics: and obviously, this places some {\it physical conditions} on the situations or systems that are being compared. 

(iii) {\it Mathematical approximations} (good or bad) to certain expressions, typically by some dimensionless number which is `large' or `small'. For example, when a series is cut-off after a certain order of interest.

And we admit that there are subtle issues about these three cases, especially about whether the limiting system or limiting properties exist, and how well they describe the actual physical system. Roughly, we have three types of linkage: idealisation, good approximation, and poor approximation.\footnote{Butterfield et al.~(2015) relate their schema to a trichotomy of Norton's (2011), allowing for both cases of idealisations and (good or bad) approximations. They relate this schema to their discussion of emergence, but this will not concern us here.} But we will not need such details here: for which we refer to e.g.~Butterfield et al.~(2015), of which we will now recall some notions:---

An approximation is a particular kind of relation between theories, as follows. Consider a theory $T_{\sm b}$ (where the subscript stands for  `better, bottom or basic') and a theory $T_{\sm t}$ (for `tainted, tangible or top' theory). Take the two theories to be related by one (or a combination of) the above approximations. Then, applying the approximation in question to the basic theory, $T_{\sm b}$, gives a new theory, $T_{\sm t}$ (related to $T_{\sm b}$ by way of idealisation or approximation). Alternatively, rather than considering the theories themselves, one considers specific phenomena (and thus specific {\it models} of the theories) and makes these approximations---the subsequent relation then still being one of idealisation or approximation. This linkage relation can be regarded as a surjective\footnote{Surjectivity is what is meant by the phrase `$T_{\sm t}$ reduces to $T_{\sm b}$ in the approximation': $T_{\sm t}$ is what is left of $T_{\sm b}$ after the approximation is made, i.e.~each element of $T_{\sm t}$ corresponds (``comes from'') at least one element of  $T_{\sm b}$.} and non-injective map between the theories, i.e.~a map: $\mbox{Approx}: T_{\sm b}\rightarrow T_{\sm t}$ (or the corresponding map between the phenomena).

The important point for scientific understanding is that the application of approximations, or linkage relations, can also be used to induce a (partial) interpretation of a theory that had none, as we now argue. And so, this is how (Approximation) comes to be a tool for interpretation.\\

Assume that $T_{\sm b}$ is a bare theory (say, a theory of quantum gravity) whose interpretation one is developing, and assume that its approximating theory, $T_{\sm t}$ (say, general relativity), already has an interpretation, i.e.~it can be mapped to the world. Furthermore, we will assume that the interpretation is ``sufficiently detailed'' that each element in the domain of application of the theory is mapped to by some element of the theory. (Though this looks like a strong condition, it is in fact innocuous: it can be done as you would expect, by restricting the domain of application of the theory such that the interpretation map is surjective). So, we have two maps: $\mbox{Approx}:T_{\sm b}\rightarrow T_{\sm t}$ is the approximative relation, or linkage, between the theories. And $i:T_{\sm t}\rightarrow D$ is the interpretation map from the top theory to (a domain in) the world. Both maps are surjective. Then these relations induce a {\it new interpretation map} for the basic theory, $T_{\sm b}$ (for some of its terms at least), and we will dub this map the {\it inherited interpretation}, $i'$. Namely, the inherited interpretation map, $i'$, results from the successive application of the two surjective maps, the approximation and the top theory's interpretation: $i':=i\circ \mbox{Approx}$, which is again surjective. The inherited interpretation is an interpretation because it maps the bottom theory to a domain of the world $D$, viz.~$i':T_{\sm b}\rightarrow D$.

An inherited interpretation of the basic theory, $T_{\sm b}$, is of course not a full-fledged interpretation of all the terms in the theory, but only a {\it starting point} for a more comprehensive interpretation. To see that more than an approximation is needed to interpret the full theory $T_{\sm b}$, consider the fact that the domain of the world, $D''$, which $T_{\sm b}$ could describe (and which an interpretation of all the terms of the theory ought to map to) may be much larger than the domain, $D$, which $T_{\sm t}$ describes (and which figures in the inherited interpretation, $i'$): e.g.~because it applies to more cases and solves more problems. In other words, the inherited interpretation, $i'$, describes a smaller or more restricted domain than the basic theory, $T_{\sm b}$, could describe. And so, what we are really after is an interpretation map to the domain $D''$, i.e.~$i'':T_{\sm b}\rightarrow D''$.\footnote{There is no claim here that strictly $D\subset D'$ (and we do not need such a strong condition). For Kuhn's incommensurability will, in the cases where it applies, imply that the full interpretation $i''$ {\it does not describe some of the elements of $D$}, which the inherited interpretation $i'$ {\it did} describe. Rather, the claim we are making is that the inherited interpretation, $i'$, can be used as a tool for developing the full interpretation, $i''$, because there is a {\it non-zero overlap}, i.e.~$D\cap D''\not=\emptyset$. This is only the claim that inconmensurability, though real, is not total: there are some elements in the world which were described by the top theory's interpretation $i$, and which are still described by $i''$. And of course, $i$ and $i''$, as interpretations, may well be inconsistent with each other (unlike $i$ and $i'$, which are consistent with each other).} Nevertheless, the inherited interpretation, $i'$, may be a good starting point for a full interpretation. This is how (Approximation) can be used to interpret a theory, from another already interpreted theory. 

This is of course how, for the most part, quantum mechanics is interpreted. For large enough systems, quantum mechanical uncertainties disappear, and abstract quantities such as the square of the wave-function can receive a straightforward interpretation (as a probability for some process to occur). Formally, this can be done by taking a limit, $\hbar\rightarrow0$ (see e.g.~Landsman (2016)). For example, one can interpret a Gaussian wave-function as describing a `delocalised particle', by considering the fact that in the limit $\hbar\rightarrow0$, the wave-function becomes better and better localised, approaching a Dirac delta function in the position variable (what we call a classical `point particle'). The interpretation of the Gaussian wave-function thus inherits its interpretation, $i'$, as {\it de}localised {\it particle} from the case in which the particle is perfectly {\it localised} (and the point particle interpretation is $i$). Ultimately, we may wish to do away completely with the concept of `particle' in quantum mechanics, and so develop an interpretation $i''$ that does not refer to them: and such interpretations are indeed suggested by, or can be guessed from, the fact that $i$ interpreted the delta function as a point particle and $i'$ interpreted the Gaussian wave-function as a delocalised particle. If no such limit were available at all, so that we did not have the sequence from $i$ to $i'$, it would be much harder to try and work out $i''$. 

It may appear from the above that the more approximations, with their inherited interpretations, one can take, the easier it is to construct a full interpretation $i''$, mapping all the physical elements of the theory to the world. But this need not be so. Think, for example, how in quantum mechanics there is a particle interpretation, which holds in a certain limit and for a certain set of examples; and a wave interpretation, which holds in other limits for other sets of examples. It is not easy to come up with an interpretation $i''$ that reconciles these two. So, different approximations may lead to mutually incompatible interpretations. 

There is of course no necessary connexion between (Approximation) and spacetime or visualisation. For example, in statistics, one may take the limit in which the number of e.g.~individuals in a population is infinite, without the limit's generating any spacetime. Thus (Approximation) is a very general linkage relation, applicable to a wide class of theories, with or without a spacetime. But, since we are here interested in the question of spacetime visualisation, we will narrow down the scope of (Approximation) (and its cousin to be discussed next, (iApproximation)), and consider only examples that {\it do} generate spaces with geometrical structures.

Notice that {\it all} theories of quantum gravity ultimately aim at reproducing general relativity in an appropriate (Approximation). And so, one can strictly speaking not say that quantum gravity theories contain no spacetimes at all. Rather, we need to distinguish between theories whose interpretation depends on their having a spacetime, and those whose interpretation is built using other tools: the latter are the theories  without a spacetime. As we will see in section \ref{nost}, the interpretive arrows, for the latter kinds of theories, can be reversed---so that new light is cast on old spacetime theories by the new theories that replace them. $T_{\sm t}$ does not help interpret $T_{\sm b}$, but the other way around. We will call this an `inverse approximative relation':

\subsubsection*{(iApproximation)~~Inverse approximative relations}
\addcontentsline{toc}{subsubsection}{(iApproximation)~~~Inverse approximative relations}

Inverse approximative relations are cases of approximations (in any of its three forms mentioned in the previous subsection: mathematical limits, comparing physical situations, mathematical aproximations) in which the bottom theory $T_{\sm b}$ is used to interpret (usually: to {\it re}interpret) the {\it top} theory, $T_{\sm t}$. The linkage relation $\mbox{Approx}:T_{\sm b}\rightarrow T_{\sm t}$ is the same as before, but in constructing the interpretation we now use one of its inverses, $\mbox{Approx}^{-1}:T_{\sm t}\rightarrow T_{\sm b}$. We say `one of its inverses' because an inverse of the linkage relation between two theories need not be unique (indeed, we cannot expect it to be unique: but one will exist, because the approximative map is surjective).\footnote{We shall not go into details here about how to choose the inverse $\mbox{Approx}^{-1}$. Suffice to say that, in the cases of interest, in which spacetime {\it emerges} out of the bottom theory, the robustness (autonomy) condition of the emergence map ensures that the interpretation is actually independent of differences of detail in the inverses. But we will not need such details.}

Now imagine that we are in the possession of some interpretation of $T_{\sm b}$, not obtained through (Approximation), but through the other tools to be introduced in this section, (Similar) or (Internal). Denote this interpretation as $i_{\tn{SI}}$ (for `similar-internal'; although which of the tools were used to construct $i_{\tn{SI}}$ is irrelevant). We obtain an inherited interpretation, $i$, of the top theory through an inverse approximative relation, defined by the successive application of the inverse of the approximative map and the interpretation map:
\bea
i:=i_{\tn{SI}}\circ \mbox{Approx}^{-1}:T_{\sm t}\rightarrow D~.
\eea

Two of the quantum gravity theories that we will consider in section \ref{nost} replace spacetime by a more fundamental discrete structure underlying it. (iApproximation) reinterprets the spacetime of general relativity in light of the interpretation of the more fundamental discrete structure which replaces spacetime. 

An analogy can be helpful here. The discovery that matter is made of atoms, which can be classified by their atomic number and electron configurations, obviously compels one to rethink whatever interpretation of matter and of its interactions one held before making that discovery. For example, one is now able to interpret chemical reactions as the rearranging of the electrons in the orbitals of the atoms involved, to break old bonds and form new ones. In a similar way, the discrete structures underlying spacetime should shed new light on old aspects of spacetime, and so help develop new interpretations. For example, some physicists may interpret the fact that the underlying structure of spacetime is discrete, and that in some cases a continuum limit cannot be taken (because the mathematical limit does not exist), as saying that continuous spacetime is a mere appearance: namely, something that looks continuous to our senses but which ought to be thought of as discrete. 

The difference between (Approximation) and (iApproximation), then, is not in their underlying formal procedures---which include the same options---but in the direction in which the interpretative arrows go. Both emphasise the fact that interpretations are {\it dynamic}: as we mentioned in section \ref{wiin}, interpretations are not unique and can change according to aims and---as we see here---in the face of new scientific theories. \\

How does (iApproximation) relate to visualisation? Even though a space with geometric structures is formally developed in the (iApproximation), this structured space is now to be reinterpreted starting from the discrete theory. So, the visualisation of the space is no longer a contingently dominant tool in (iApproximation): rather, the spacetime visualisation will now be {\it reinterpreted} in the light of the basic theory, which (in the cases of interest for our aims) cannot be so visualised. 

Notice that, in practice, (Approximation) and (iApproximation) will often appear combined, in a hermeneutic circle, which can be described as follows. One starts with a tentative interpretation, $i_0$, of the basic theory, $T_{\sm b}$. This interpretation could have been obtained either directly through the tools (Similar) and (Internal), or via (Approximation), from the spacetime interpretation of the top theory, $T_{\sm t}$---general relativity, say. One then improves the interpretation, $i_0$, of the basic theory using (Similar) and (Internal). These tools lead to new interpretative aspects which are added to the tentative interpretation, $i_0$. At the same time, one may try to strip this interpretation, as much as possible, from any of the original spacetime elements that it may still have. The result of this is a new interpretation, $i_1$, of the basic theory. In turn, this interpretation may have consequences for, and thus call for a revision of, one's original interpretation of the top theory, $T_{\sm t}$: specifically, it may call for a revision of the interpretation of spacetime in general relativity. Thus one constructs an inherited interpretation of $T_{\sm t}$, viz.~$i_2:=i_1\circ \mbox{Approx}^{-1}$, which should give an improved, or more fundamental, interpretation of the spacetime in general relativity (or any other spacetime theory that one started with). 

An interpretation that is obtained through (Similar) and (Internal), but without the use of (Approximation), may well be rather minimal. And then it may well be desirable to get a fuller interpretation through the use of (Approximation), e.g.~to compare with the spacetime of general relativity. But, in such a case, we still insist that the two interpretative steps are conceptually distinct: and that even before one uses visualisation, one already can have a (minimal) interpretation of the theory. This illustrates our point, that spacetime is not strictly needed for interpretation. 

Strictly speaking, (iApproximation) is not a tool for interpreting theories  without a spacetime: but a tool for reinterpreting general relativity (or other spacetime theories) from the basic theories  without a spacetime underlying them. And so, if we are correct, it predicts that a good theory of quantum gravity would lead to a reinterpretation of general relativity's spacetime interpretation. We will focus less on this tool, but we will indicate where the authors we discuss make use of (iApproximation).

\subsection*{(Similar) Similarities in the use of concepts between theories}
\addcontentsline{toc}{subsection}{(Similar)~~~~~\,Similarities in the use of concepts between theories}

The first tool just discussed, (Approximation), concerned a relation of parenthood between theories, i.e.~a one-way entailment. The second tool concerns a relation of {\it similarity between different theories}. Similarities between theories can indeed suggest ways of interpreting the concepts or terms in a theory. Similarities will usually relate individual terms between theories, and how these are related to the overall theoretical structure of a theory.\footnote{All the tools discussed in this paper can be used for theories, for theory parts, and for instantiations, representations or exemplifications of theories (which in De Haro (2016) and De Haro and Butterfield (2017) are called `models', though not in the model-theoretic sense). In what follows, what we say about theories holds for these other cases as well: but we will not say so explicitly, and we will simply use the word `theories'.}

Notice that similarity between different theories comprises a spectrum of likeness, ranging from (almost) identity to analogy.\footnote{The `almost' here is because when we compare {\it different theories}, as we do in (Similar), there will always be some disanalogies, somewhere in the interpretation of the theoretical structure. There will never be a rule that automatically tells us, from the interpretation of an old theory, how a new theory ought to be interpreted. One should always look out for surprises!} Thus (Similar) will admit the same varieties.

The similarities between two interpreted theories can be either in the bare theory, i.e.~in the formalism, or in the interpretation. As follows:

(i)~~{\it Formal similarity:} a similarity between parts of the mathematical formalism of two theories can be used to draw consequences about the interpretations of certain terms in the bare theory (mathematical formulas, concepts, etc.). That is, a given formal similarity between two theories is used to construct a ``matching'' interpretation for one of the two theories. 

An example of formal similarity is that of conservation laws in Maxwell's theory, general relativity, and quantum mechanics (and admittedly, the similarity is not {\it only} formal, but also interpretative!). The electrical charge and current in Maxwell's theory play the role of sources for the electric field, and they are associated with a conservation law that follows as an identity from Maxwell's equations. This general idea (of being a source for a field, and of being associated with a conservation law which expresses a certain identity that follows from the equations of motion) can then be applied to other theories, like general relativity, where the stress-energy tensor plays a very similar role: of a source, which satisfies a conservation law, which follows from Einstein's equation. In quantum mechanics, the probability density and the probability current play similar roles as the charge density and current in Maxwell's theory, and obey a formally similar conservation law, which follows from the Schr\"odinger equation: but there are also important disanalogies between quantum mechanics and Maxwell's theory, which Max Born emphasised. Nevertheless, the formal similarity (in this example: a case of analogy) was historically of instrumental value in developing the interpretation of the wave-function, e.g.~through the analogy between charge conservation and the law that was interpreted as a law of conservation of probability (and that analogy is, in fact, the starting point of the development of the de Broglie-Bohm theory). 

(ii)~~{\it Interpretative similarity:} a similarity between some aspects of the interpretations of two theories can be used to construct a fuller, or more complete, interpretation of one of the theories, matching the interpretation of the other theory. 

Often, interpretative similarity also helps unravel a formal similarity, so that technical-mathematical progress can be made. 't Hooft's planar diagrams, which in section \ref{detail} will be combined with (Approximation), will be an example of this. Thus formal and interpretative similarity usually go hand-in-hand.

Interpretative similarity also involves cases in which the interpretation of terms within a theory is derived directly from observation or experiment. However, this will not be our focus in this paper.

As an example of interpretative similarity, consider the interpretation of the formalism of quantum mechanics, which is a case of (almost) identity: the Born rule receives the same interpretation across many quantum mechanical theories and models (from elementary quantum mechanics to quantum field theory to quantum gravity),\footnote{The Born rule is of course part of the interpretative core of elementary quantum mechanics. However, applications of quantum theory in new domains, such as quantum field theory and quantum gravity, rely on (Similar), in that the correct identification and application of the Borne rule is not straightforward. The details of the theoretical structure, however, differ: e.g.~probabilities may have to be appropriately renormalised, and the relevant parameters on which the probability depends need to be identified, since they may differ from elementary quantum mechanics. And so, it is in the formal relations between the Schr\"odinger equation, the wave-function, and the absolute value squared of the wave-function, that a specific quantity can be interpreted as `the probability (density) of finding the system in state such and such', or as `the probability of finding the value of a certain quantity between such and such values'. In other words, probabilities are interpreted on analogy with the case of elementary quantum mechanics, using the intertheoretic relations with other quantities in the theory.} as do the eigenvalues and eigenvectors of Hermitian operators. 

In section \ref{nost}, we will encounter examples of both formal and interpretative similarities. 

\subsection*{(Internal) Internal criteria}
\addcontentsline{toc}{subsection}{(Internal)~~~~\,Internal criteria}

The third tool concerns a {\it single theory of model}. More precisely, it concerns {\it internal criteria} which can be used to interpret the concepts or terms in the bare theory. As in the case of (Similar), these criteria can vary: from structural similarity, to analogy. Also, as in the case of (Similar), the internal criteria can be of two types: formal or interpretative similarity, both internally to a single theory. Formal or mathematical similarity between two parts of the theory is used to draw consequences about the interpretations of certain expressions. Interpretative similarity is used to interpret uninterpreted terms in the theory, or to further develop an existing interpretation.

For example, in quantum mechanics, the fact that certain quantities (such as the standard deviation of a distinguished operator, divided by its time change) are formally conjugate to the energy, in the sense that they satisfy Heisenberg's uncertainty principle together with the energy, has led some to take these quantities as measures for the uncertainty in time.\footnote{Of course, a theorem by Pauli proves that there is no self-adjoint operator representing time, if the Hamiltonian is bounded from below. But the intended interpretation is only that a specific quantity can be taken for uncertainty in time, not that time is described by a self-adjoint operator, or that the Hamiltonian is bounded from below.} The reasoning here is from structural similarity within the theory of quantum mechanics itself: if position and momentum satisfy certain structural relations, and energy and some other quantity satisfy the same structural relations, then one is entitled to take this other quantity as a measure for the uncertainty of time (which is the expected quantity to satisfy those structural relations). 

(Internal) will often be applied {\it after} external methods like (Approximation) or (Similar) have been applied. Once some terms in a theory receive a tentative interpretation through external methods, internal comparison can lead to the interpretation of other, still uninterpreted, terms in the theoretical structure. For example, this is how non-observational quantities (quantities which are not observable according to current methods, or quantities which will perhaps always remain out of reach) can receive an interpretation: formally, from their relation with other quantities which already have an interpretation. This is the case with for example dark matter and dark energy, which can currently not be observed directly (and it is a possibility that they never will) but which were postulated in order to explain other phenomena. Comparing the rotation curves of galaxies to the predictions of Newton's theory, the missing dark matter, if it exists, can be interpreted as interacting gravitationally, and its mass can be calculated.

Internal criteria are the correct interpretative criteria for certain kinds of dualities, as first argued in Dieks et al.~(2015) and, more specifically within the current framework for interpretation, in De Haro (2016:~\S1.3).\footnote{For the kind of heuristic work that dualities do in theory construction, see De Haro (2018).} Equivalent theories are to be seen as `models (in the sense of instantiations) of the same theory', and an interpretation of a theory can be worked out which starts from the (minimally interpreted) `common core', or shared structure, between the models. Though they involve different models, such interpretations are cases of (Internal) not (Similar), because the criteria of interpretation are internal to the theory. \\

From the three tools discussed above, (Approximation) is the one which gives rise to visualisation of theories in terms of space and time. Visualisation is of course not the only way to use this tool. In the next section, we will study how visualisation is used in quantum gravity theories, before we turn (in section \ref{nost}) to cases in which there is no spacetime and no visualisation.

We should add here a warning about the analysis of the examples that we will carry out in sections \ref{viseff} and \ref{nost}. We will be concerned with details about the question whether a particular theory does or does not employ spacetime for its interpretation. But such verdicts should be seen as answering the question that motivates this paper---viz.~whether understanding of theories without spacetime can be had---and it should not be taken as a judgment, on our side, whether a particular research programme is more or less feasible, comes more or less close to producing a ``good'' theory of quantum gravity, etc. Our aim here is to get at the notion of understanding through interpreting, and not appraising quantum gravity theories.

\section{Visualisation via effective spacetimes}\label{viseff}

In this section, we discuss theories in which (Approximation) is used as a tool to develop a {\it visual} interpretation, i.e.~an interpretation that employs space (and perhaps time). Thus, in the use of (Approximation) we will consider here, either a new spacetime appears in the theory, or an already existing spacetime is transformed in some way that is useful for interpreting the theory, thus rendering it intelligible (in the sense of section \ref{wiin}). 

We will first introduce two distinctions that broadly characterise the effective spacetimes that we will be dealing with:\\

(a)~~{\it Effective spacetimes: `visualisable' does not entail `physically real'.} The general question, of how visualisation in spacetime is used to construct an interpretation, is logically independent from the question whether physicists consider a spacetime to be physically real and-or observable: hence our term, `effective' spacetimes. Visualisation helps rendering a theory intelligible, even if the space is not considered to be real. For example, in the method of image charges in electrostatics, a boundary condition (such as the vanishing of the electric potential on a plane bounding the problem) is replaced by a (set of) image charge(s). This is a visual method (an interpretation) because it allows to picture the otherwise rather abstract boundary condition imposed on the potential (and, more importantly, it allows to do calculations in an intuitive way, which would otherwise be rather more cumbersome). But of course, the imaginary charges that are visualised at the other side of the plane are not physically real. Rather, we compare our world with another {\it possible world} which is much easier to visualise, and in which for each charge a mirror charge of opposite sign can be found. In the cases we discuss in this section, it is {\it spacetime itself} that may not be physically real, but auxiliary---like the imaginary charges.

Such auxiliary, or effective, spacetimes are allowed in our interpretation: for, as we remarked in section \ref{wiin}, interpretation maps not only map to the real world, but often also to other possible worlds---as in the case of the imaginary charge, which is not real, yet its use hinges on its interpretation and properties as a `charge'. In the example at hand, this interpretation is an intension, because our comparison with another possible world in which there are imaginary charges does not need to consider contextual details, but only idealised details about the charges.\\ 

(b)~~{\it Physical vs.~merely mathematical spaces.} But despite the above remark, that visualisation is logically independent of the question of physical reality: a second distinction needs to be made. Namely, whether a space is interpreted as {\it physical} (in the actual world, or in some possible world) or as merely {\it mathematical}, is an important question, which we will address. So, when we refer to `theories without spacetime', we have in mind the former meaning of spacetime (for more on the distinction, see section \ref{mms}). This is because it is very hard (and it would be contrived) to formulate a theory that does not make any use at all of the mathematical concept of space, i.e.~of a set with some added structure. For example, the numerical parameters of physical theories all take values in fields, which are examples of spaces. As soon as we have a quantity taking values in $\mathbb{R}$, we have a space. It would surely be unreasonable to demand, on the grounds that we must have a theory without spacetime, that $\mathbb{R}$ may not appear anywhere in our theories---at the very least, such a revisionist project is not the kind of project that quantum gravity researchers have in mind when they talk about theories `without a spacetime'. So, in what follows, our concept of space will be physical not merely mathematical, and when we talk about visualisation we will mean visualisation through a space with physical properties (such as particles, fields or forces defined on it). And we submit that visualisation through physical spaces (whether in our world, or in another possible world) contributes to the intelligibility of the theory more than visualisation via merely mathematical spaces, because in the former case physicists can use physical arguments and intuitions, in addition to mathematical ones.

Agreed, the distinction between merely mathematical and physical spaces can sometimes be blurry, if the mathematical spaces come with sufficient structure (as, say, normed vector spaces, or even inner product spaces). For such a mathematical space may, if appearing within a physical theory, easily inherit a physical interpretation (in the way described in section \ref{3tools}). But this need not always be so: for example, the geometric interpretation of phase space in classical mechanics, as a symplectic manifold, is not physical: for the phase space is itself not deemed to be physically real, under the standard interpretation of classical mechanics. And yet it provides a visualisation which helps physicists to apply concepts such as the existence of forms on this space, the independence of the choice of coordinates, etc., which helps them better grasp certain aspects of the theory (such as e.g.~canonical transformations). But in this paper we will not be concerned with such cases of mathematical but not physical spaces, since what is at stake in theories of quantum gravity is of course spacetime itself, and how theories without a physical spacetime can be interpreted. \\

Summarising this discussion, (a)-(b): the topic of visualisation in scientific understanding should in general also consider spaces which are mathematical and not physical. But for our specific question, of how theories without spacetime can be interpreted, only the question of whether there is a {\it physical} space is relevant (second fork:~(b))---but such a space need not be physically real (first fork:~(a)). \\

Our first case study, of 't Hooft's theory of planar diagrams (section \ref{thooftp}), will be a theory in which one starts with a spacetime of one kind (a space in which particles move), and ends up, after a suitable (Approximation), with a spacetime of another kind (a space in which stringlike bits of matter rather than point particles propagate). In the second example (section \ref{MM}), we will start with a theory in which there is no spacetime, but in which a curved spacetime is developed after a similar kind of limit. These examples are closely connected to the question of the `emergence of spacetime', i.e.~whether spacetime can appear as a novel and robust entity in a theory in which there is no spacetime (or only a very simple one). While we will not address the question of emergence here, it should be clear that we are considering cases in which physicists and philosophers alike do use the term (see De Haro (2018a, 2018b), Crowther (2016), Rickles (2013)).

\subsection{'t Hooft's planar diagrams}\label{thooftp}

In this subsection, we work out our first example, of visualisation via Feynman-like diagrams (actually called 't Hooft diagrams), in Yang-Mills theory (for more details, see De Haro (2018b)).

\subsubsection{Brief summary of the idea}

Let us here explain the idea in simple terms. (More details are in \S\ref{detail}.) The idea is that $T_{\sm b}$ is the theory of the strong interactions (Yang-Mills theory) and $T_{\sm t}$ is a theory whose objects are strings. And the relation between the two theories involves taking an (Approximation) of $T_{\sm b}$ (defined in \S\ref{detail}), the so-called `'t Hooft approximation'. The approximative theory is $T_{\sm t}$, i.e.~there is a map $\mbox{Approx}: T_{\sm b}\rightarrow T_{\sm t}$.

In the 't Hooft approximation, the Feynman diagram expansion of $T_{\sm b}$ can be reorganised into the string diagram expansion of the Nambu-Goto string (via so-called 't Hooft diagrams), i.e.~$T_{\sm t}$: the dominant diagrams which contribute in the 't Hooft approximation are the so-called `planar diagrams', i.e.~those that can be drawn on a surface of genus zero (the plane). In this way, the Feynman diagram expansion of $T_{\sm b}$, instead of being classified by the number of loops (as in the Feynman diagram expansion for point particles in $T_{\sm b}$), is now classified, in $T_{\sm t}$, by the number of holes and boundaries of the stringlike surface: so $T_{\sm t}$'s expansion is an expansion in terms of the topology of stringlike surfaces, as in the case in string theory.

The original interpretation of these strings was as an explanation of hadronic phenomenology, and in particular 't Hooft was looking for an explanation of quark confinement. The idea that strings held quarks together seemed to provide a reasonable explanation for this. 

As we remarked in the preamble to this section, the fact that these strings cannot be literally interpreted as holding quarks together is besides the point. For we are not concerned here with the literal physical truth of the constructions (cf.~point (a) in Section 1.1), but with how they lead to interpretations that render the theory intelligible. Again, the interpretation is an intension: it abstracts away from the (technically involved!) details of accelerator experiments, and just looks at the interactions between the particles, as described by the basic theory.

As we will see in the next subsection, the visual recognition of the stringlike surfaces in the 't Hooft approximation, and the ensuing topological expansion of the amplitudes, is a very useful tool for gaining insight into expressions that would have been very hard to handle using the conventional methods.

\subsubsection{The 't Hooft approximation, in more detail}\label{detail}

The basic theory $T_{\sm b}$ considered  by 't Hooft (1974) is the theory of the strong interactions, a Yang-Mills theory with gauge group $\mbox{U}(N)$, containing quarks and gluons. He made the remarkable observation that, in the limit of large $N$ (keeping the combination $\l:=g_{\tn{YM}}^2\,N$ fixed, i.e.~also sending $g_{\tn{YM}}\rightarrow0$, where $g_{\tn{YM}}$ is the gauge coupling of the Yang-Mills theory), the theory reduces to a theory of string-like surfaces, with only planar diagrams contributing, as explained above. That is, $T_{\sm t}$ is best interpreted as a string theory.

This remarkable result was obtained through relatively simple physical methods: first, a slightly different notation for the Feyman diagrams, which use double lines, rather than wiggly lines, for the gluon propagators; quarks are denoted by single lines. The double lines then give the visual impression of spanning `surfaces', and are interpreted as such, given that the formulas also carry the same structural similarities to theories of surfaces. Second, Euler's topological formula 
\bea\label{Euler}
V-E+F=2-2g-b
\eea
is applied to the Feynman diagrams. Here, $V$ is the number of vertices, $E$ the number of edges, and $F$ the number of faces, on any given triangulation of the diagram. $g$, on the right-hand side, is the genus of the surface on which the diagram is drawn, and $b$ is the number of boundaries (the number of quark loops). 


Euler's formula marks the transition from thinking in terms of single and double lines, to thinking in terms of surfaces: a transition from a theory of point particles, sweeping out world-lines, to two-dimensional surfaces with smooth interactions. The string interpretation of $T_{\sm t}$ thus developed is a combination of (Approximation) (which provides the initial interpretation in terms of Feynman diagrams, which are then reinterpreted as thickened diagrams), and the (Similar)-ity of the diagrams and expressions obtained in the approximative theory $T_{\sm t}$ with string theories. 

In terms of this new interpretation, it was easier to characterise the theory obtained in the limit: as a theory in which only planar diagrams (those that can be drawn on the plane) contribute. 

But 't Hooft went further: the new interpretation allowed him to make predictions {\it away from} the strict $N=\infty$ limit. Namely, the original Feynman diagram expansion of the amplitudes of the theory can now be better and more significantly reorganised as a topological genus expansion, akin to the ones found in string theory: 
\bea\label{thooft}
{\cal A}(\l,N)&=&\sum_{g,b=0}^\infty N^{2-2g-b}~A_{g,b}(\l)\\
A_{g,b}(\l)&:=&\sum_{n=0}^\infty c_{n,g,b}~\l^n~.\nonumber
\eea
where $g$ is the genus of the surface (the number of holes) and $b$ the number of boundaries. The interpretation of $n$ is as the number of gluon loops. Notice that $\l=g^2_{\tn{YM}}\,N$, so the expansion in $\l$ brings extra powers of $N$; however, since $\l$ is held fixed, the above is the way to organise the expansion in the 't Hooft approximation: in powers of $N$. 

Notice that, for each power of $N$ in \eq{thooft}, an infinite number of Feynman diagrams (corresponding to the infinite summation over $n$, for fixed $g,b$) contribute to $A(\l)$. Thus 't Hooft's expansion is a very efficient reorganisation of the Feynman diagrams, since it picks out the ones that contribute at each order in $N$ (which are an infinite number). 

This example falls within the class discussed in the first interpretative tool, (Approximation), in section \ref{3tools}. Namely, the string interpretation, $i$, of $T_{\sm t}$ now introduces a reinterpretation, $i':=i\circ\mbox{Approx}$, of the Yang-Mills theory, the basic theory, $T_{\sm b}$. Thanks to the relation between $T_{\sm b}$ and the theory of strings, $T_{\sm t}$, in the 't Hooft approximation, we can reinterpret (a certain sector of) Yang-Mills theory itself as `containing strings'. These sorts of considerations have recently led more generally to relations between gauge theories and theories of gravity.

Notice the close connexion between (Approximation) and visualisation in this example. The Feynman diagram expansion, the double line notation, and Euler's topological formula Eq.~\eq{Euler}, were essential in developing the string interpretation of the gauge theory. So, this example instantiates the discussion in the preamble of this section, about starting with a space of one kind, in $T_{\sm b}$ (the one-dimensional worldline of a particle), and ending up with a space of a different kind, in $T_{\sm t}$ (the two-dimensional worldsheet of a string). 

That this interpretation leads to the {\it intelligibility} of certain aspects of gauge theory can be seen from the further history of the subject. 't Hooft used his approximation in work in which he obtains further results about such gauge theories, which he could not have obtained with conventional Yang-Mills methods. Also, 't Hooft's paper has received many other applications since (the paper has more than 4,000 citations): it has led, for example, to some deep developments in string theory, namely the 't Hooft approximation underlies the relation between gauge theories and string theories known as gauge-gravity duality.

\subsection{Random matrix models}\label{MM}

In the previous subsection, we discussed how the application of the 't Hooft approximation allows physicists to reinterpret a Yang-Mills theory as a theory of strings, rather than point particles; such that in the 't Hooft approximation, the leading contribution to the scattering amplitudes is given by planar diagrams. Furthermore, the exact amplitude can be completely reorganised as a topological expansion for stringlike surfaces. Thus the string interpretation gives insight into otherwise intractable calculations. The theory starts with an initial spacetime picture, and develops {\it another} spacetime picture in the approximation.

In this section, we will give an example of a model which does not start off with a spacetime, but which in the 't Hooft approximation does give rise to a curved two-dimensional surface with holes, viz.~a Riemann surface. For an introduction to random matrix models aimed at philosophers, and a discussion of emergence, see De Haro (2018b).

\subsubsection{Brief summary of the idea}\label{summm}

The physical idea is now as follows. It is actually a simplified setting for what 't Hooft had in mind: the basic quantities are still non-abelian {\it matrices} (like in Yang-Mills theory), but without any space or time dependence, and consequently with no spacetime dynamics---there is no kinetic term in the Lagrangian: only a potential term. In this way, one gets what is called a `random matrix model', i.e.~a model in which the fields are plain $N\times N$ matrices, distributed with an appropriate probability density (given by the Lagrangian) and no spacetime dependence. 

Because of symmetry, the physical quantities do not depend on all $N^2$ components of the matrix, but only on its $N$ eigenvalues. And so, the dynamics of this model is the dynamics of the eigenvalues, subject to a potential. In principle, these eigenvalues can take any complex values, and so they correspond to $N$ points on the complex plane. However, in the saddle-point approximation, the potential imposes a ``zero-force condition'', and the eigenvalues tend to fill up parts of the complex plane (physicists speak of ``condensation of the eigenvalues''), forming branch cuts. One can show that, in the 't Hooft limit, this `plane with branch cuts' is best described as a Riemann surface with non-trivial cycles: some cycles are around the branch cuts, and others connect different branch cuts. 

In other words: one started with a theory in which one had only matrices and no spacetime. Then, in the 't Hooft approximation, a non-trivial geometry emerges out of the eigenvalue condensation. It is this geometry which is extremely useful for the intelligibility the dynamics of the random matrix model. And as before, all physical quantities have a topological expansion like Eq.~\eq{thooft}.

The (Approximation) allows visualisation, because it fabricates a two-dimensional Euclidean space (there is no time) in which the theory is easy to visualise, and therefore easier to interpret. The difference with 't Hooft's original example is that the random matrix model itself does not have an initial spacetime: it is an algebraic structure. Thus, the Riemann surface that one ends up with in the 't Hooft limit is a {\it novel geometric structure}, within a theory which had no initial spacetime. 

The physical interpretation of this Riemann surface can be taken much further, vindicating 't Hooft's claim that large-$N$ gauge theories really are string theories. The Riemann surface turns out to be best interpreted as embedded in a 6-dimensional Calabi-Yau manifold, on which a closed topological string theory is defined. This is because the original random matrix model described an open topological string theory. The cycles of the Riemann surfaces on the closed string side then correspond to projections of the 3-cycles of the Calabi-Yau, and the quantities in Eq.~\eq{thooft} can be interpreted as quantities (especially, the free energy) of the closed topological string theory. The string-theoretic picture of course adds to the usefulness of the 't Hooft approximation; in particular, the higher-dimensional geometry makes vivid its use in visualisation.

\subsubsection{Random matrix models, in more detail}

The basic quantity of the model is a (real or complex) $N\times N$ matrix, $\F$, sometimes with additional restrictions (usually: self-adjoint). The matrix $\F$ contains $N^2$ independent numbers, and there is neither space nor time dependence. In addition, there is a potential function $W(\F)$ which is a polynomial of degree $n+1$ in $\F$ (and hence it is a matrix; there is no kinetic term, because there is no spacetime). The basic quantity of the model is the partition function, as follows:
\bea
Z={1\over\mbox{Vol}(\mbox{U}(N))}\int\dd\F~\exp\left(-{1\over g}\,\Tr\, W(\F)\right),
\eea
where Tr indicates the trace of the potential function (which is a matrix). $g$ is the coupling constant of the potential. Of course, there is some arbitrariness here, corresponding to the possibility of changing $\F$ by unitary transformations, $U\in\mbox{U}(N)$,  $\F\mapsto U\,\F\,U^{-1}$, which leave the potential invariant. The measure is also $\mbox{U}(N)$-invariant. Thus one divides by the (infinite) volume of $\mbox{U}(N)$. 

Since all expressions are invariant under unitary transformations, the matrix $\F$ can be diagonalised, and the integral reduces to an integral over the $N$ eigenvalues of $\F$, $\l_1,\ldots,\l_N$:
\bea
Z&=&\int\prod_{i=1}^N\dd\l_i~\D(\l)^2~\exp\left(-{1\over g^2}\,\sum_{i=1}^NW(\l_i)\right)\nn
\D(\l)&:=&\prod_{1\leq i<j\leq N}(\l_i-\l_j)~,
\eea
where $\D(\l)$ is a factor coming from the integration measure. This contribution can be added in the exponential, using the logarithm.

The above integrals can be evaluated using the saddle-point approximation. The terms in the exponential (potential plus the logarithm of the measure term) are extremised, as follows:
\bea\label{wprime}
{1\over g^2}~W'(\l_i)=2\sum_{j\not=i}{1\over\l_i-\l_j}~.
\eea
This is a set of $N$ equations for the eigenvalues, which are the independent variables of the theory. This equation can be interpreted as the classical equation of motion of the random matrix model. The term on the right-hand side, coming from the measure factor, induces a Coulomb-type repulsion between the eigenvalues. 

As in the previous subsection, the 't Hooft approximation takes $N$ to be large while keeping $g^2N$ fixed. One also introduces a continuous function $\l$, defined by: $\l_i=\sqrt{N}\,\l(i/N)$ (in practical calculations, one takes the $N\rightarrow\infty$ limit of the expressions involved: but this is not required). The density of eigenvalues, $\r(\l)={1\over N}\sum_{i=1}^N\d(\l-\l_i)$, becomes a function on the real axis, normalised to $\int\dd\l\,\r(\l)=1$. The eigenvalues then fill domains of the real axis; these domains are continuous, but may have several disconnected components. 

In the continuum limit, the discrete set of $N$ equations, Eq.~\eq{wprime}, is replaced by a continuous equation for $\l$, where the summation is replaced by an integral. However, it is more useful to first multiply Eq.~\eq{wprime} by a factor of $1/(\l_i-x)$, and to sum over $i$. Here, once can roughly think of $x$ as the value of (the trace of) the quantum field $\F$ (not necessarily satisfying the equation of motion Eq.~\eq{wprime}).

The result of the manipulations described in the previous paragraph, starting from Eq.~\eq{wprime}, is the following equation:
\bea
y^2=W'(x)^2+f(x)~.
\eea
$y(x)$ is defined in terms of the derivative of the potential, $W'(x)$ (a polynomial of degree $n$ in $x$), and $f(x)$, which is a polynomial of degree $n-1$. $x$ can be thought of as the trace of the quantum field $\F$. Notice that both $x$ and $y$ are complex.

The above equation defines a hyperelliptic curve in the $(x,y)$-plane, because the right-hand side is a polynomial of degree $2n$ with distinct roots. Such a hyperellictive curve defines a Riemann surface of genus $n-1$, as claimed in \S\ref{summm}. This Riemann surface, roughly speaking, provides the location of the eigenvalues, given the configuration of the matrix which is dominant in the saddle-point approximation: since the eigenvalues now form a continuum along the $n-1$ handles of the Riemann surface. 

Now that a geometric interpretation of the random matrix model, as a Riemann surface with $n-1$ handles, has been obtained, it also follows that the amplitudes are of the type Eq.~\eq{thooft}, with $g=n-1$ and (in this case) $b=0$, i.e.~all the terms with boundaries vanish. Clearly, the geometric interpretation gives a reoganisation of the amplitudes, like before, which allows a different interpretation of the random matrix model in terms of surfaces.\\

The interpretative lesson is as follows: (Approximation), along with (Similar) and (Internal), functions here as a tool like in the previous subsection, because it allows the development of a visual picture---a geometric interpretation in terms of a Riemann surface---which gives insight into otherwise intractable calculations. It thereby allows one to perform those calculations. The difference with the previous example is that here we do not start with a field living on a spacetime; rather, the spacetime picture only manifests itself after the 't Hooft approximation is taken. 

\section{No spacetime and no visualisation}\label{nost}

Our two examples in section \ref{viseff} focused on interpretations which are constructed by developing an effective spacetime in a specific (Approximation). This effective spacetime then allows physicists to visually interpret the theory. In this section, we move on to cases in which visualisation and effective spacetimes are not needed, and not used as interpretative tools---even if spaces might be present somewhere else in the theoretical structure. Rather, interpretation is developed through the use of the tools (Similar) and (Internal) from section \ref{3tools}. 

We will first give, in section \ref{mms}, our generic conception of a `theory without a spacetime'. Our notion of `theory without spacetime' will not be absolute, but relative to the models which are used for the interpretation. On such a conception, a theory without spacetime will be a theory that does not employ (for the interpretation in question) a physical space, though it may use a mathematical space---and we sketch a simple conception of this notion in \S\ref{mms}, and physical spaces might be present somewhere in the theory. Then, in the next three sections, we will move on to our examples: of causal sets (\S\ref{causet}), loop quantum gravity (\S\ref{lqg}), and group field theory (\S\ref{gft}). 

\subsection{Merely mathematical spaces and theories without spacetime}\label{mms}

In this subsection, we collect our remarks (a)-(b), from the preamble of section \ref{viseff}, to work towards a conception of a `theory without spacetime'---which we will give at the end of the subsection; as well as the related notion of a `merely mathematical space'.

Recall our two distinctions, from the preamble of section \ref{viseff}, which bear on the kinds of spacetimes used for interpretation: \\

(a) an {\it auxiliary spacetime} with visualisation (i.e.~a merely effective spacetime) vs.~a {\it physically real} spacetime; 

(b) {\it physical space(time)} vs.~{\it mathematical space}. \\

For the project of this section, we do not want to allow interpretations entailing spaces that are `physical', either in a real physical sense (e.g.~the space of general relativity), or in a fictitious, but still physical sense (e.g.~phase space). Therefore, in this section we focus on examples that do not contain any of the spaces of the disjunction (a), and contain only the {\it mathematical spaces} of the disjunction (b). We will dub such spaces `merely mathematical', and we give a more explicit conception of them below.\\

But we first make two warnings about merely mathematical spaces, in relation to the tools (Similar)-(Internal) which we will illustrate in the next three subsections. The first warning is about the tools to be used in theories without spacetime; the second is about visualisation of merely mathematical spaces:---\\

(1) {\it (Similar) and (Internal) can be had in theories without spacetime, despite the presence of a spacetime somewhere in the theory.} This is because our claim, that we are dealing with `theories without spacetime', does not entail the absolute non-existence of any sort of physical space anywhere in the theories and models we will present. Ultimately, {\it every} realistic theory of quantum gravity attempts to recover the pseudo-Riemannian spacetime of general relativity, typically in an (Approximation) or an (iApproximation): and so, every theory without spacetime, attempting to describe the world, must at one point or another develop a physical space. But remember what our no-spacetime claim refers to: not to the {\it presence}, but to the {\it use} of spacetime as a tool for interpretation. More specifically, the claim is that these theories do not use (Approximation) when they develop their approximations; and that these theories do illustrate the use of (Similar) and (Internal) without reference to spacetime or visualisation. They also sometimes illustrate (iApproximation), with its inverse relation between bottom theory and spacetime visualisation. So, the title of this section---`no spacetime and no visualisation'---should be read as relative to the tools which the interpretation of the theory actually uses.\\

(2) {\it Merely mathematical spaces can be free from spacetime visualisation.} We should say a bit more about our distinctions (a)-(b) recalled above, between physical and merely mathematical spaces, even if we do not attempt at a systematic treatment of this question here (nor will we need it: for a discussion of a concrete case, cf.~De Haro (2018b:~\S4.2)). Indeed, between a finite set of numbers, a set of points with the cardinality of the continuum, and a manifold endowed with a metric, there is an almost continuous spectrum of possibilities for what we can call `merely mathematical' vs.~`physical structure'. Whether a space qualifies as `merely mathematical' or as `physical' depends on how strongly it is physically interpreted. Accordingly, we will call these two cases `merely mathematical' vs.~`physical' spaces. And of course, even in the case of a set consisting of three elements, one can use visualisation, viz.~three encircled dots. But it would be foolish to mistake such a visualisation for a counterexample to the claim that we are dealing with a theory without spacetime. For, first: such a visualisation is not needed. And second: the picture of the encircled dots is rather more a mnemonic, or a visual way to fix ideas, than providing a spacetime interpretation of a set consisting of three elements. It is not {\it spacetime visualisation}.\\

With these two comments in mind, and motivated by our notion of visualisation, we now settle for the following notion of a {\it merely mathematical space}, viz.~as either: (A) a discrete (finite or infinite) space, or (B) a continuous space, but with no rich geometric structures on it (e.g.~only topological structure).\footnote{Oriti calls such discrete structures `pre-geometric, non-spatiotemporal structures', `atoms of space', cf.~Oriti (2016). Notice that we are being very restrictive here about the kinds of spaces that we allow as merely mathematical. It is possible to weaken our requirement to include spaces with geometric structures, but stripped of their geometric interpretations. However, we will not need such generalisations, and so we will not consider them.} 

{\it Theories without spacetime} will consequently, for our purposes in this section, be theories that use merely mathematical spaces, and no physical spaces, for interpretation: indeed, physical spaces are not used in the specific interpretations that we will review from the literature. 

How will merely mathematical spaces figure in theories without spacetime, and how do they relate to the physical space which emerges in (Approximation)? Remember that the merely mathematical spaces are either discrete, or continuous but without geometric structure.  In the three cases we will study, such spaces will constitute part of the basic structure of the theory (e.g.~they go into the fundamental degrees of freedom which are then quantised) but they are pre-geometric, in that they have no straightforward geometric interpretation. These spaces will underlie the physical, pseudo-Riemannian space of general relativity, which will be recovered as an (iApproximation). But crucially, as we stressed in point (1) of section \ref{mms}, physicists do {\it not} use (Approximation) to develop key parts of the interpretation. Thus, the relation between the merely mathematical space and the physical space will be that the former underlies---in fact, replaces---the latter; and that the latter is derived from and emerges from the former, in an (iApproximation). Physical space is derived, explained and, ultimately, interpreted, from merely mathematical space plus an (iApproximation): and not the other way around.

\subsection{Causal Sets}\label{causet}

\begin{figure}
\begin{center}
\includegraphics[height=4cm]{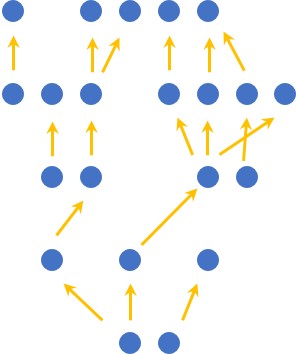}
\caption{A Hasse diagram of a causet.}
\label{hasse}
\end{center}
\end{figure}

Our first example is causal set theory. A causal set (or `causet')\footnote{In our presentation here, we follow W\"uthrich (2012). See also Sorkin (1997, 2003) and Fey (2005).} is an ordered pair $(C,\leq)$, consisting of a set $C$ of elementary events, and a relation $\leq$ defined on $C$. The relation $\leq$ is a `partial ordering' (i.e.~it is reflexive, antisymmetric, and transitive: but such details will not concern us here), as well as locally finitary (roughly, the number of events `within some interval' of the causet is finite). A causal set can be represented by a Hasse diagram, as in Figure \ref{hasse}, where the partial ordering $\leq$ is indicated by the arrows. 

So, causets stipulate a discrete structure underlying what we know as the spacetime of general relativity, and a fundamental {\it causal} relation $\leq$ which does not assume a time ordering (rather, temporal ordering is supervenient on causal relations). Neither space nor time are present in this theory; the set of points and the relations in Figure \ref{hasse} are not {\it in} spacetime; rather, they are the elements {\it out of which} spacetime will be constructed. It is this `out of which', or `underlying', relation, between the causet and spacetime, that is important: because most causets are {\it not} interpreted as spacetimes---the number of causets vastly outnumbers the number of spacetimes. 
Thus it is not spacetime that is fundamental, but events and causal relations. 

The diagram's interpretation is primarily motivated by notions of causality and logic (cf.~also the quotes from Sorkin, at the end of this section). And one might argue that the notion of causality is best understood in connexion with temporal becoming. However, this need not be so, because causality can also be modelled on logical relations of entailment. At any rate, the causal set theory does not itself propose an account of what it means by causation: causation is a primitive term, part of the definition of the theory. And so, one is entitled to resist the temptation of reading time (or spacetime) into a diagram like Figure \ref{hasse}. It is not spacetime which does here the explaining: it is the explanandum. 

Despite the previous remark, one may still be inclined to see Figure \ref{hasse} as a sort of ``discretisation'' of spacetime. And, to a certain extent, one is entitled to such an picture, heuristically at least. The problem is that it does not seem to square with the logic behind causets. For not all causets can be embedded in a spacetime, and so it is not the case that all causets are discretisations of spacetimes. Rather, the causets are the more fundamental structures underlying spacetime. 

How does a causet `underlie' a spacetime? The idea is that the causet can be embedded (under some technical conditions) into a relativistic spacetime (and also, it is hoped that such embedding is somehow unique, i.e.~that the same causet cannot be embedded into two different spacetimes, at least at scales above the discreteness scale, W\"uthrich (2012:~\S2.2)). 

In fact, one of the problems of causets is that the vast majority of them do not seem to allow for such embeddings, and so do not give rise to spacetimes. This is not surprising, since no `dynamics' has been specified. So far, we have not specified a theory that dictates which sets $C$ are physically admissible, and which causal relations are defined. We will not go into details here---suffice it to say that a dynamics can be produced, roughly, by `assigning probabilities' to causets. And here, there are two basic approaches: one approach starts with a given manifold and then discretises it through a method called `sprinkling', i.e.~taking a sufficiently large number of points distributed at random so as to have a unique embedding. But other approaches are much more bottom-up, and select the configurations of points via a microscopic principle---something like an action that allows the formulation of a path integral.\\

{\it Hermeneutic circles.} At the end of our discussion of (Approximation) and (iApproximation) in section \ref{3tools}, we mentioned that one useful way to understand the process of developing interpretations is as a `hermeneutic circle', in which scientists start with a {\it tentative} interpretation, $i_0$ of $T_{\sm b}$---for the causet, $i_0$ could e.g.~be a combination of the logical notions of partial ordering that motivate the definition of a causet, and perhaps something like ``thinking of a Hasse diagram as a discrete spacetime diagram''. That is, the tentative approximation is not yet without spacetime. Using (iApproximation), one can then develop an interpretation of $T_{\sm t}$ (general relativity, say) based on the causet. Having developed a new interpretation of $T_{\sm t}$ based on $T_{\sm b}$, one can then go back to $T_{\sm b}$ and {\it drop} the parts of its tentative interpretation, $i_0$, that are not essential, to build a more adequate interpretation: for instance, its spacetime connotations could be dropped, so that one now works out a better interpretation of the causet in terms of causal relations or logic. The hermeneutic circle here functions as an {\it interpretative circle in which we strip the interpretation of, or purify it from, its spacetime connotations}: resulting in an interpretation that is more in line with the fundamental character of the theory, which serves as a foundation for spacetime theories, rather than it being interpreted according to them.\\

We now show explicitly where causal sets use tools that do not require visualisation, i.e.~where they use (Similar) and (Internal) as interpretative tools. Below we give four examples of interpretations which use (Similar), i.e.~they compare to other theories, and also use (Internal).\\

(1) {\it The sequential growth dynamics interpretation.} Rideout et al.~(2000:~\S2) develop a classical sequential growth dynamics for causal sets, for which they introduce a new terminology (see also interpretation (2), below). They call the dynamics in question `transitive percollation', which is a simple instance of a sequential growth dynamics---a random evolution of the causet. And they give a `static perspective' on transitive percolation, by presenting an algorithm for generating a random poset, without requiring a notion of `evolution': only elements, probabilities, and relations. In the same paper (\S6), they also use ideas from Ising models from condensed matter physics to `illustrate how non-gravitational matter can arise dynamically from the causal set without having to be built in at the fundamental level' (p.~1). \\

(2) {\it The family interpretation.} In the same paper, Rideout et al.~(2000:~\S1.1) introduce yet another metaphor to explain their ideas: they introduce a `parent-child' terminology to explain the process of growth: 

`The causal sets which can be formed by adjoining a single maximal element to a given causet will be called collectively a {\it family}. The causet from which they come is their {\it parent}, and they are {\it siblings} of each other. Each one is a {\it child} of the parent. The child formed by adjoining an element which is to the future of every element of the parent will be called the {\it timid child}. The child formed by adjoining an element which is spacelike to every other element will be called the {\it gregarious child}.'

We have quoted the whole passage to show how differently a causet is now interpreted. Of course, genealogical descendancy relations are familiar from posets in the mathematical literature. But it is interesting that Rideout et al.~here choose not to read the causet as spatiotemporal, but in terms of human relationships: even adding human traits, such as `timid' and `gregarious', that are {\it not} part of the usual poset interpretation. And, though such interpretations are obviously not physical interpretations, they do help physicists fix ideas about how to use the theory, and hence they fall squarely within what we call an `interpretation', in section \ref{wiin}. 
Thus this is another case of (Similar). 

This parent-child interpretation is not, within the context of Rideout et al.~(2000), a mere curiosity: it is a central part of the interpretive framework, in section 1 of the paper. We submit that it is simply a reflection of the difficulties of working theories where there are no space and time. The authors did not hesitate to use a pedagogical analogy with human relationships when this seemed to suit their aim. To be sure, this interpretation is thrown in with several more technical concepts from sequential growth dynamics, such as transitive percolation (a simple instance of sequential growth dynamics including a single probability rule) and the Markov sum rule (the requirement that the probabilities add up to one). This illustrates our point, from section \ref{wiin}, that a single interpretation may not suffice to fulfill all the goals. 

The human interpretation also reinforces our point that, even though some interpretations are more prone to be seen as having some spatiotemporal connotations, even if logically speaking that need not be so, some other intepretations do not have that connotation at all.

Also, Rideout et al.~give a reinterpretation of their results in terms of spins living on the causet (\S6), which helps the reader get further insight into the calculations---indeed the main calculations in the paper can now be reproduced using the spin picture. All of the above are cases of (Similar), combined with other (Internal) arguments which further refine the interpretation.\\

(3) {\it The logic interpretation.} Our point, that a spacetime interpretation is not needed, has perhaps been best made by Sorkin (2010), in the title of the following paper: ``Logic is to the quantum as geometry is to gravity''. Clearly, the message is that `geometrical' pictures, used to interpret gravity, are to be replaced by `logical' pictures, in the case of quantum gravity, since there is no spacetime. This reflects well the logical motivation that is behind the entire causal set theory. The paper indeed emphasises the logical interpretation of the causets and the logical motivation of the causet programme (\S2): the space of all events is structurally a Boolean algebra, with connectives {\it or, and, not,} and {\it xor}. He also defines coevents as the duals of events (i.e.~functions assigning truth values to events). These notions amount to what he calls an `anhomorphic logic', a logical reinterpretation of the sum over histories approach to quantum mechanics: ``For a scientist, logical inference is---or I believe it should be---a special case of dynamics... the logic I'm speaking of concerns physical events, not strings of words'' (\S5). Now also the causal set dynamics finds a logical reinterpretation. Clearly, for the reader familiar with logic and Boolean algebras, these topics can be used as tools, in the way of (Similar).\\

(4) {\it Quantum field-theoretic interpretations.} Sorkin (2017) uses a construction of quantum field theory that does not require the use of the equation of motion. Since the equation of motion is not needed, he can then apply this construction to a causal set (where the equations of motion of the fields do not hold exactly, but only approximately). Thus he is able to borrow familiar concepts from quantum field theory: Green's function, Wightman function, vacuum, entropy, inner products of fields, etc. These notions already have interpretations in quantum field theory that can be stripped of any spacetime connotations, which do not apply for the causet. For example, a Green's function can be interpreted as an impulse response: a notion that is used for a wide range of dynamical systems, as responses (outputs of a certain kind) to input signals, e.g.~in signal processing. Like before, these notions no longer refer to a time ordering as in quantum field theory, but rather to the causal or logical order that is appropriate for the causet. This `generic' interpretation (output response to input signal) indeed suffices to be able to use the notions effectively. (Internal) can then be used to add more layers to the interpretation, specific to the fields of the causet. Clearly, the other notions (Wightman function, vacuum, entropy, inner products of fields) can be given similar interpretations which, in their use here, are free from spacetime. In conclusion, quantum field theory is a framework so familiar that it allows physicists to work with causets---even if there is no underlying spacetime!\\

The four interpretative strategies that we have reviewed above are cases of (Similar), because they apply interpretations from other fields to causets. Furthermore, none of these interpretations strictly require spacetime---rather, they are what underlies the spacetime interpretation of general relativity (the quantum field theoretic interpretation is perhaps one which has some time connotations, but as mentioned---it is an interpretation stripped of its time connotations). And as we mentioned in section \ref{3tools}, (Internal) criteria are automatically used as well. For instance, in the parental relations considered above, the relation of being a {\it sibling}, or being a {\it gregarious child}, simply follows from the other relations (parenthood and childhood) already defined. This is an application of (Internal). 

\subsection{Loop Quantum Gravity and Spin Foams}\label{lqg}

Our second example is loop quantum gravity (and its more recent, associated development: spin foams), an approach which arose from the canonical quantisation programme of quantum gravity, but which has given rise to an entire subfield of quantum gravity.\footnote{Due to space constraints, we cannot give details here about even the most elementary aspects of the theory. We refer the interested reader to Rovelli and Vidotto (2015).}

We will here illlustrate two aspects about loop quantum gravity:\footnote{We refer to `loop quantum gravity and spin foam theory', collectively, as `loop quantum gravity'.} (a) Its being a theory without spacetime, under the conception of `theory without spacetime' introduced in section \ref{mms}. (b) How to develop an interpretation of the theory, physicists use (Similar) and (Internal).\footnote{Remember our remark (1) in section \ref{mms}, that even if a spacetime must be obtained in an (Approximation), this is no objection to the development of already a simple interpretation of the theory. This is because, as we said in the last paragraph of section \ref{mms}, the explanatory relation between merely mathematical spaces and physical spaces is reversed in these theories: it is the former that explain the latter.}

The first point---loop quantum gravity's being a theory without spacetime---is already illustrated, for example, in the titles of some of the sections of Rovelli and Vidotto (2015): ``The end of space and time'' (\S1.2), ``Fuzziness: disappearance of classical space and time'' (\S1.4.2), ``Physics without time'' (\S2). Also explicitly in their wording (which actually also emphasises the topic of understanding): 

``The description of spacetime as a (pseudo-) Riemannian manifold cannot survive quantum gravity. We have to learn a new language for describing the world: a language which is neither that of standard field theory on flat spacetime, nor that of Riemannian geometry. We have to understand what quantum space and what quantum time are. This is the difficult side of quantum gravity, but also the source of its beauty.'' (p.~19).

This verbal description is illustrated technically in the section immediately following (\S1.3), about quantised geometry. And here, as in our remark (1) in section \ref{mms}, we should bear in mind the fact that the programme is one of replacing the classical geometry of general relativity by a discrete structure---a merely mathematical space---which underlies physical space. 

The merely mathematical space that lies at the basis of the development of loop quantum gravity is a tetrahedron. A tetrahedron can be parametrised by a set of four vectors, $\{{\bf L}_a\}$, or $\{L^i_a\}$, one for each face of the tetrahedron (where the index $a=1,\ldots,4$ labels each of the faces, and $i=1,2,3$ are spatial indices). Since this is a quantum theory, the vectors $L^i_a$ are quantised: they are taken to satisfy the angular momentum algebra of SU(2). The areas and volumes constructed from these basic objects thus come out quantised. 

Notice that spacetime is to be explained and derived from these objects, and not the other way around. In fact, it seems that the limit of an infinite number of quanta does not even exist. In any case, the geometric intuition is of little help here: for these are discrete quanta of space, fuzzy objects whose values cannot be determined simultaneously. And so, (Approximation) is here not the primary aid.

Another way to say this is as follows. The classical picture one starts with is one of a piece of space with a tetrahedron. This picture is only used to identify for the variables to quantise (since the theory is supposed to reproduce general relativity in the right (Approximation)). But the interpretation of the quantum theory does not start from the classical interpretation; it starts somewhere else, namely from the quantum theory itself,\footnote{This is then another case of a hermeneutic circle. It is also similar to what happens in the interpretation of dualities (De Haro (2015), Dieks et al.~(2015)). Though two dual models have initial classical interpretations (called external interpretations), the discovery of duality, especially as a quantum relation, prompts one to develop a new interpretation (an internal interpretation), which starts from the duality itself. In the case of gauge-gravity dualities, the internal interpretation drops most of the geometric, and even the topological, structures present in the external interpretation, as was shown in detail in De Haro (2015, 2016).} as the rest of the section makes clear:---

A useful tool to be used here is (Similar): Rovelli and Vidotto (2015) exploit the analogy with quantum mechanics. The algebra of the $L^i_a$'s is a slight generalisation of the usual SU(2) angular momentum algebra, familiar from quantum mechanics:
\bea\label{algebrall}
{}[L^i_a,L^j_b]=i\,\e_{ab}\,\ell^2\,\e^{ij}{}_k\,L^k_a~,
\eea
and so the problem can be treated mathematically using the ordinary SU(2) algebra.\footnote{The above algebra is isomorphic to the direct products of copies SU(2) algebras, one for each pair of variables.} Indeed, most of the exercises given in this chapter as practice for the reader are about SU(2), spin, the Pauli matrices, and the angular momentum variables. They are {\it not} about spacetime or general relativity. It is evident that the authors are drawing on the reader's knowledge of quantum mechanics to become familiar with this case, rather than from visualisation in spacetime.

(Internal) is also used, for example, in discovering the role of Newton's constant in this theory. Once the algebra, Eq.~\eq{algebrall}, has been related to the angular momentum algebra of quantum mechanics, an interpretation can be developed for the constant $\ell^2=8\pi\g\,\ell^2_{\tn{Pl}}=\g\,{8\pi G\hbar\over c^3}$, as a `quantum of area', since the area (and volume) operators will come quantised in units of $\ell$. At this particular point, a comparison with general relativity {\it is} being used, in order to interpret the eigenvalues of these operators as areas and volumes.\footnote{For this reason, loop quantum gravity seems more tied with a classical spacetime interpretation than do causal sets or group field theory.} Nevertheless, it seems that we have here a case of the hermeneutic circle mentioned in sections \ref{3tools} and \ref{causet}: an initial interpretation as `area' or `volume' is corrected, in (iApproximation), by the appearance of an underlying discrete algebra that replaces what we usually call areas and volumes. Areas and volumes are no longer to be interpreted as classical geometric quantities, but rather as classical limits of operators with a discrete spectrum, and in terms of states that can be in superpositions of eigenvalues. 

(Similar) is also used to develop interpretations that borrow from theories other than elementary quantum mechanics: quantum field theory techniques are used in \S1.4. The interpretation emphasises the phase transitions that might occur, and that these are independent of the microscopic cut-off. ``The system is characterised by a physical and {\it finite} cut-off scale---the atomic scale---and there are no modes of the [iron] bar beyond this scale. The bar can be described as a system with a large but {\it finite} number of degrees of freedom. The short-distance cut off in the modes is not a mathematical trick for removing infinities, nor a way for hiding unknown physics: it is a genuine physical feature of the system. Quantum gravity is similar: the Planck-scale cut-off is a genuine physical feature of the system formed by quantum spacetime.'' (p.~31). \\


An alternative, more modern, way to construct these pre-geometric structures are the so-called {\it spin foams}. The construction is similar to that of loop quantum gravity, but now the dual links of a triangle on the plane (rather than a tetrahedron) are quantised. Again, at this point this is a merely mathematical space that is going to be quantised. The Hilbert space is then $L^2$ on the SU(2) associated with the rotations about each link, up to gauge transformations (rotations about the nodes). The transition amplitudes constructed from concatenating multiple copies of these objects in an approate way (using the Feynman path integral), form a spin foam. 

Oriti (2013:~\S III.A) emphasises the fact that the Hilbert space built as above does not contain {\it geometric} data, but rather ``discrete combinatorial structures (the graphs), labelled by algebraic data only.'' 

This point is also emphasised by Markopoulou: ``These approaches start with an underlying microscopic theory of quantum systems in which no reference to a spatiotemporal geometry is to be found.'' (p.~129). As well as the fact that mathematical spaces can also have other, non-spatiotemporal interpretations (as we saw, for causets, in section \ref{causet}), obtained via (Similar)-ity. For example, there is a formal similarity with circuit models of quantum computation which can, in some sense, be used: ``While [the graph] has the same properties as a causal set, i.e.~the discrete analog of a Lorentzian spacetime, it does not have to be one. For example, in the circuit model of quantum computation, a circuit, that is, a collection of gates and wires also has the same properties as [the graph] and simply represents a sequence of information transfer which may or may not be connected to spatiotemporal motions.'' (in Oriti , p.~136).

The tools available here are again (Similar) and (Internal). And as before, mathematical structures and techniques, familiar from quantum mechanics and quantum field theory, are used effectively to develop intuition and peform calculations. Hence the (Similar)-ity with quantum mechanics and quantum field theory are used. Once these structures are understood, (Internal) is used to shed light on other parts of the structure.

Among the mathematical techniques used from quantum mechanics and quantum field theory, are: harmonic analysis on the group SU(2), the Haar measure on the group, the Clebsch-Gordan coefficients, Wigner's 3j- and 6j-symbols, the representation theory of groups, spinors. To get the classical (Approximation), coherent states are obtained.

For example, A.~Perez, ``The spin foam representation of loop quantum gravity'' explicitly illustrates the analogies, which amount to the use of (Similar): 

``A sum over gauge-histories in a way which is technically analogous to a standard path integral in quantum mechanics. The physical interpretation is however quite different... The spin foam representation arises naturally as the path integral representation of the field theoretical analog of $P$ [a projector operator] in the context of loop quantum gravity.'' (p.~275). 

\subsection{Group Field Theory}\label{gft}

Our third example, of group field theory, seems to be the one which makes the least reference to any kind of spacetime. Indeed, while causal sets and loop quantum gravity both start with a merely mathematical space which can be seen as a {\it discretisation} of the physical spacetime, group field theory does not do this: its merely mathematical space consists of fields.\footnote{For example, the spacetime dimension $d$ is encoded in the number of fields, and so it is not a topological quantity.} 

Group field theory, as Oriti defines it, is a quantum field theory over a group manifold. This manifold is not a physical spacetime but a merely mathematical space: a group. The dynamics for the fields is given by an action, and the Hilbert space is defined as the space of square integrable functions on the group manifold, up to gauge transformations. 

A Fock space for this theory can then be defined in the usual way, by taking an infinite sum over multiple copies of the Hilbert space. The states of many quanta in this Fock space are spin networks (enjoying a specific graph representation), and so the Fock space is the space of network vertices. Oriti emphasises the fact that the vacuum on which the theory is based has no topological or geometric information, and so neither does the Fock space. The Feynman diagrams derived from group field theory perturbation theory are related to the spin foams of section \ref{lqg} (from which group field theory originally arose). Also here, there is no straightforward continuum field interpretation, since the continuum limit is problematic. 

To explain group field theory, Oriti uses (Internal) in a way similar to the spin foams discussed in the previous sections. That is, he emphasises the quantum field-theoretic and group-theoretic aspects of group field theory, which are familiar to the theoretical physicist:

``There are two main ways of understanding GFTs. The first is as a second quantised field theory of spin network vertices, each corresponding to a quantum of the field $\phi$ and labelled by the $d$ group or Lie algebra elements,\footnote{$d$ is the spacetime dimension.} constructed in such a way that its quantum states are generic superpositions of spin networks and its Feynman diagrams are spin foams.'' (Oriti (2013:~\S III.B)).

The above use of (Similar) includes appeal to quantum field theory, Lie algebras and Feynman diagrams are here quoted as theories with which the reader is already familiar, and which are of help in interpreting group field theory. 

In the same article, Oriti makes another use of (Similar), by comparing group field theory with related (and older) theories in quantum gravity, namely loop quantum gravity and spin foams. Namely, he reinterprets group field theories as a second quantisation of simplicial geometry, thus giving a discrete version of gravity (Oriti 2013:~\S III.B).\\


{\it Comparison between group field theory and random matrix models.} Group field theory shares close historical ties with the random matrix models discussed in section \ref{MM}: for, in a certain sense, group field theory is a generalisation of random matrix models. The main difference between the interpretation of random matrix models (as used in quantum field theory) and that of group field theory is as follows. For random matrix models, the explanandum is the quantum field theory itself (i.e.~the random matrix model, or the quantum field theory of which it is part): for random matrix models, {\it spacetime is the explanans}. And thus, random matrix models develop {\it spacetime interpretations} using (Approximation), which can shed light on the random matrix models. Group field theory, on the other hand, has a different aim, and so do its interpretations. One central aim of group field theory is to explain the spacetime of general relativity, and so {\it the spacetime is the explanandum, while group field theory is the explanans}. Consequently, group field theory does not use (Approximation), except---perhaps: cf.~the paragraph just below---as a temporary stage of a hermeneutic circle (cf.~section \ref{3tools} and \ref{causet}).

In the case of group field theory, we have not identified the presence of a hermeneutic circle. Group field theory seems to start with purely algebraic objects, and it never seems to use spacetime concepts: except, of course, after spacetime is made to emerge.\footnote{The only point where a certain spacetime connotation could perhaps be found is in the use of Feynman diagrams. But these are not spatiotemporal.} And so, it does not seem subject to the hermeneutic circle that causets and loop quantum gravity do often seem to require, in which the theory's interpretation is stripped of its spacetime connotations, so that a more fundamental interpretation can be made to underlie the construction of spacetime. Of course, this may not be the last word on the subject: such a hermeneutic circle may well be required: for, as we emphasised, {\it all} approaches to quantum gravity aim at reproducing the spacetime of general relativity. 

\section{Conclusion}\label{conclusion}

Theories of quantum gravity in which there is no spacetime pose an obvious question: how should such theories be interpreted? How can one make sense of them, i.e.~how can such theories be rendered intelligible?

To answer this question, one needs an account of what an interpretation of a scientific theory is, and of how interpretation relates to scientific understanding. Our preferred account of interpretation, in terms of suitable maps that are intensions or extensions, was argued to be suitable for the task. An interpretation was argued to be a precondition for the intelligibility of the theory. Intelligibility is, in turn, a necessary condition for achieving understanding of the phenomena, according to the contextual theory of understanding that we have adopted in this paper. Depending on the context, different conceptual tools may be used for constructing interpretations that render a theory intelligible.

We classified three conceptual tools which physicists have at their disposal to interpret theories, with or without a spacetime. (Approximation) relates two theories through an approximative scheme, and thus develops an interpretation. (Similar) draws on similarities (formal, or conceptual) with other already known theories. (Internal) derives its interpretation from the theory itself. 

We gave several examples to illustrate how these tools are used in actual practice in quantum gravity. 't Hooft's approximation was used to develop a new spacetime within the theory (whether the theory already had a spacetime, or none), and this spacetime was used to (re-)interpret the theory. On the other hand, causal sets, loop quantum gravity, spin foams, and group field theory, developed interpretations that did not need an appeal to a spacetime (or, in some cases, had no spacetime at all). Nevertheless, one has to admit that all these approaches retain traces of spacetime {\it connotations} in their interpretations: which illustrates the fact that visualisation, though not necessary, is a contingently dominant tool. We characterised the development of such interpretations as a `hermeneutic circle', in which interpretations are stripped, where possible and required, of those connotations.


\section*{Acknowledgements}
\addcontentsline{toc}{section}{Acknowledgements}

We thank Daniele Oriti for a helpful discussion of some of the materials contained in this paper. We also thank Jeremy Butterfield, Dennis Dieks, and two anonymous referees for comments on the draft. SDH thanks Silvia De Bianchi, Carl Hoefer, and the audience at the Department of Philosophy of the Universitat Aut\`onoma de Barcelona. SDH was supported by the Tarner scholarship in Philosophy of Science and History of Ideas, held at Trinity College, Cambridge.

\section*{References}
\addcontentsline{toc}{section}{References}

Butterfield, J.N.~and Bouatta, N.~(2015). ``On Emergence in Gauge Theories at the 't Hooft Limit''. {\it European Journal for Philosophy of Science}, 5 (1), pp.~55-87.\\
\\
Carnap, R. (1947) {\em Meaning and Necessity}, Chicago: University of Chicago Press\\
\\
Crowther, K.~(2016). {\it Effective Spacetime}. Springer International Publishing Switzerland.\\
\\
De Haro, S.~(2015). `Dualities and emergent gravity: Gauge/gravity duality'. {\em Studies in History and Philosophy of Modern Physics}, 59, 2017, pp.~109-125. \\
\\
De Haro, S.~(2016). ``Spacetime and Physical Equivalence''. To appear in {\it Space and Time after Quantum Gravity}, Huggett, N.~and W\"uthrich, C.~(Eds.).\\
http://philsci-archive.pitt.edu/13243.\\
\\
De Haro, S.~and Butterfield, J.N.~(2017). `A Schema for Duality, Illustrated by Bosonization'. In: Kouneiher, J.~(Ed.), {\it Foundations of Mathematics and Physics one century after Hilbert}. Springer. http://philsci-archive.pitt.edu/13229.\\
\\
De Haro, S.~(2018). ``The Heuristic Function of Duality'. {\it Synthese,}\\ https://doi.org/10.1007/s11229-018-1708-9.  arXiv:1801.09095 [physics.hist-ph].\\
\\
De Haro, S.~(2018a). `Towards a Theory of Emergence for the Physical Sciences', Forthcoming in the {\it European Journal for Philosophy of Science.}  http://philsci-archive.pitt.edu/16256. \\
\\
De Haro, S.~(2018b). ``The Emergence of Space, Illustrated by Random Matrix Models''. In preparation.\\
\\
De Haro, S.~and De Regt, H.~W.~(2018). ``A Precipice Below Which Lies Absurdity? Theories without Spacetime and Scientific Understanding''. {\it Synthese,}\\
https://doi.org/10.1007/s11229-018-1874-9. arXiv:1807.02639 [physics.hist-ph].\\
\\
De Regt, H.~W.~(2014). ``Visualization as a tool for understanding'', {\it Perspectives on Science}, 22, pp. 377-396.\\
\\
De Regt, H.~W.~(2017). {\it Understanding Scientific Understanding}. Oxford University Press.\\
\\
De Regt, H.~W., and Dieks, D.~(2005). ``A contextual approach to scientific understanding''. {\it Synthese}, 144 (1), 137-170.\\
\\
Dieks, D., Dongen, J. van, Haro, S. de~(2015), ``Emergence in Holographic Scenarios for Gravity'', {\it Studies in History and Philosophy of Modern Physics} 52 (B), 203-216. \\
\\
Dowker, F.~(2005). ``Causal sets and the deep structure of spacetime,''
  doi:10.1142/9789812700988-0016
  gr-qc/0508109.\\
\\
Faye, J.~(2014). {\it The Nature of Scientific Thinking---On Interpretation, Explanation, and Understanding}. Palgrave MacMillan.\\
\\
Frege, G. (1892), `\"{U}ber Sinn und Bedeutung', {\em Zeitschrift f\"{u}r Philosophie und philosophische Kritik}, pp.~25-50; translated as `On Sense and reference', in P.T. Geach and M. Black eds. (1960), {\em Translations from the Philosophical Writings of Gottlob Frege}, Oxford: Blackwell.\\
\\
Hoefer, C.~and Smeenk, C.~(2016). ``Philosophy of the physical sciences''. In: Paul Humphreys (Ed.), {\it The Oxford Handbook of Philosophy of Science}. Oxford University Press. Pp. 115-136.\\
\\
Huggett, N.~and W\"uthrich, C.~(2013). ``Emergent spacetime and empirical (in)coherence'', {\it Studies in History and Philosophy of Modern Physics} 44(3), 276-285.\\
\\
Lewis, D. (1970), `General semantics', {\em Synthese}, 22, pp.~18-67; reprinted in his {\em Philosophical Papers: volume 1}  (1983), Oxford University Press.\\
\\
Markopoulou (2009), ``New directions in background independent Quantum Gravity'', in {\it Approaches to Quantum Gravity}, Oriti, D.~(Ed.), pp.~129-166. CUP: Cambridge.\\
\\
Norton, J.~D.~(2011). ``Approximation and Idealization: Why the Difference Matters'', {\it Philosophy of Science}, 79 (2), pp.~207-232.\\
\\
Oriti, D.~(Ed.), 2009, {\it Approaches to Quantum Gravity}. CUP: Cambridge.\\
\\
Oriti, D.~(2013). ``Disappearance and emergence of space and time in quantum gravity'', {\it Studies in History and Philosophy of Modern Physics} 46, pp.~186-199.\\
\\
Oriti, D.~(2016). ``Space and time are emergent, in quantum gravity. What is cosmology, then?'' Talk given at the conference {\it Foundations of Physics}, LSE, London. 16/07/2016.\\
\\
Perez, A.~(2009). ``The spin foam representation of loop quantum gravity'', in {\it Approaches to Quantum Gravity}, Oriti, D.~(Ed.), pp.~272-289. CUP: Cambridge.\\
\\
Rickles, D.~(2013). ``Duality and the emergence of spacetime''. {\it Studies in History and Philosophy of Modern Physics}, 44 (3), pp.~312-320.\\
\\
Rideout, D.~P.~and Sorkin, R.~D.~(2000). ``A Classical sequential growth dynamics for causal sets,''
  {\it Physical Review D}, 61, ~p.~024002,   doi:10.1103/PhysRevD.61.024002  [gr-qc/9904062].\\
\\
Rovelli, C.~and Vidotto, F.~(2015). ``Covariant Loop Quantum Gravity. An Elementary Introduction to Quantum Gravity and Spinfoam Theory'', CUP: Cambridge. \\
\\
Ruetsche, Laura (2011). {\it Interpreting Quantum Theories}. Oxford University Press.\\
\\
Sorkin, R.D.~(1997). ``Forks in the road, on the way to quantum gravity,''
{\it International Journal of Theoretical Physics}, 36, p.~2759.
  doi:10.1007/BF02435709
  [gr-qc/9706002].\\
\\
Sorkin, R.D.~(2003). ``Causal sets: Discrete gravity,''
  doi:10.1007/0-387-24992-3-7
  gr-qc/0309009.\\
\\
Sorkin, R.~D.~(2010). ``Logic is to the quantum as geometry is to gravity,''
  arXiv:1004.1226 [quant-ph].\\
\\
Sorkin, R.~D.~(2017). ``From Green Function to Quantum Field,''   arXiv:1703.00610 [gr-qc].\\
\\
't Hooft, G.~(1974). ``A Planar Diagram Theory for the Strong Interactions''. {\it Nuclear Physics}, B72, pp.~461-473.\\
\\
Van Fraassen, B.C.~(1989). {\it Laws and Symmetry}. Clarendon Press.\\
\\
Van Fraassen, B.~C.~and Sigman, J.~(1993). {\it Interpretation in Science and in the Arts}, in G.~Levin (ed.), {\it Realism and Representation}. University of Wiscounsin Press.\\
\\
W\"uthrich, C.~(2012). ``The structure of causal sets''. {\it Journal for general philosophy of science}, 43 (2), pp.~223-241.

\end{document}